\documentclass{article} 
\usepackage{iclr2027_conference,times}

\usepackage{amsmath,amsfonts,bm}

\def\eqref#1{equation~\ref{#1}}

\def\1{\bm{1}}

\DeclareMathAlphabet{\mathsfit}{\encodingdefault}{\sfdefault}{m}{sl}
\SetMathAlphabet{\mathsfit}{bold}{\encodingdefault}{\sfdefault}{bx}{n}

\usepackage[table,dvipsnames]{xcolor}
\definecolor{MyBlue}{RGB}{0,172,238}
\definecolor{MyPink}{RGB}{237,0,139}
\usepackage[
  colorlinks=true,
  linkcolor=red,
  citecolor=MyBlue,
  urlcolor=MyPink
]{hyperref}
\usepackage{url}
\usepackage{pifont}
\usepackage{amssymb}
\usepackage{graphicx}
\usepackage{booktabs}
\usepackage{multirow}
\usepackage{algorithm}
\usepackage{algpseudocode}
\usepackage{colortbl}
\usepackage{caption}
\usepackage{wrapfig}
\usepackage{enumitem}

\usepackage{float}
\usepackage{tabularx}
\usepackage{array}

\usepackage{makecell}
\usepackage{colortbl}
\usepackage{subcaption}
\definecolor{groupteal}{RGB}{220,239,238}
\definecolor{promptboxdata}{RGB}{76,142,188}    
\definecolor{promptboxtrain}{RGB}{107,165,120}  
\definecolor{promptboxbench}{RGB}{186,120,122}  
\newcommand{\styleref}[1]{\textcolor{red}{#1}}

\usepackage[most]{tcolorbox}
\tcbuselibrary{skins}
\newcommand{\newpromptbox}[2]{%
  \newtcolorbox{#1}[1]{%
    enhanced,
    title={##1},
    colbacktitle=#2,
    coltitle=white,
    colback=#2!6,
    colframe=#2,
    fonttitle=\bfseries,
    boxrule=0.6pt,
    sharp corners,
    breakable,
    shadow={2pt}{-2pt}{0mm}{black!20},
  }%
}
\newpromptbox{promptboxdata}{promptboxdata}
\newpromptbox{promptboxtrain}{promptboxtrain}
\newpromptbox{promptboxbench}{promptboxbench}

\title{From Surface to Depth: \\Towards Cognitive Appraisal Reasoning \\in Multimodal Emotion Understanding}
\author{
Jia Li$^{1}$, Yichao He$^{1}$, Yangchen Yu$^{1}$, Qiankun Li$^{2}$, Xinyi Li$^{1}$, Baiyi Ye$^{1}$\\
\textbf{ Zhenzhen Hu$^{1}$, Richang Hong$^{1}$, Erik Cambria$^{2}$}\\
$^{1}$Hefei University of Technology\hspace{1mm}$^{2}$Nanyang Technological University\\
\texttt{jiali@hfut.edu.cn, cambria@ntu.edu.sg}
}

\iclrfinalcopy 
\begin{document}

\maketitle
\lhead{}

\begin{abstract}
Recent multimodal large language models (MLLMs) increasingly incorporate explainable reasoning for emotion understanding. However, reasoning based mainly on observable affective cues can reduce emotion understanding to superficial cue-label associations, giving rise to the \textit{Clever Hans effect}\footnote{\textbf{Clever Hans Effect:} A system exhibits seemingly intelligent behavior but actually relies on hidden cues, data biases, or superficial associations rather than genuinely understanding the mechanisms underlying the task.}. Such shortcuts become unreliable when affective cues are implicit, conflicting across modalities, linguistically misleading, or obscured by redundant details. 
In contrast, human emotions are shaped by how individuals interpret and evaluate surrounding events beyond observable cues. Inspired by appraisal theories of emotion, we formulate multimodal emotion understanding as a progression from perception to cognitive appraisal, and introduce a dataset, a model, and a benchmark to support this novel paradigm. \ding{182} \textbf{CogEmo-40K} is a large-scale instruction-tuning dataset constructed through a perception-to-appraisal pipeline to elicit evidence-grounded reasoning across six cognitive appraisal dimensions underlying emotion.
\ding{183} \textbf{CogEmo-MoE} is a compact sparse MLLM that introduces interleaved MoE blocks for appraisal-specific adaptation, enabling effective appraisal reasoning at a substantially smaller scale than typical emotion MLLMs.
\ding{184} \textbf{CogEmo-Bench} introduces an Appraisal Evidence Quality Score (AEQS) to assess cognitive-affective understanding across six complementary appraisal dimensions, addressing the limitation of conventional emotion metrics that evaluate what emotion is predicted but not why it arises. 
Extensive experiments show that our paradigm not only leads CogEmo-Bench (CogEmo-MoE: \textbf{70.60} vs. 56.30 AEQS for GPT-5.2; \textbf{71.43\%} vs. 63.10\% accuracy and \textbf{57.17\%} vs. 52.19\% M-F1 for Qwen3-Omni), but also exhibits strong cross-domain generalization.  Our findings suggest that the perception-to-appraisal reasoning can move beyond surface-level cue-label associations toward more reliable multimodal emotion understanding and closer cognitive alignment between MLLMs and humans. Project Page: 
\url{https://github.com/MSA-LMC/CogEmo}
\end{abstract}

\section{Introduction}

Emotion understanding is not simply a matter of recognizing observable affective expressions. A prominent perspective in emotion research holds that emotional responses depend not only on what happens, but also on how individuals interpret and evaluate what happens to them~\citep{ellsworth2003appraisal,moors2013appraisal}. The same event may lead to different emotional responses depending on whether it supports or obstructs personal goals, whether the situation is perceived as manageable, who is considered responsible, and how future consequences are anticipated~\citep{roseman1991appraisal,yeo2024associations,liang2026emotions}. From this perspective, observable behavior provides only part of the information needed for emotion understanding: perception reveals \textit{\textbf{what is happening}}, whereas cognitive appraisal concerns \textit{\textbf{what the situation means to the individual}}~\citep{ong2015affective,skerry2015neural,teotia2026evaluating}.

In recent years, multimodal large language models (MLLMs) have advanced emotion recognition toward more flexible and interpretable multimodal emotion understanding by generating natural-language explanations alongside emotion predictions~\citep{lian2025affectgpt,cheng2024emotion,zhao2025r1,zhang2026mersurvey}. However, plausible explanations do not necessarily imply genuine reasoning about why an emotion arises~\citep{luo2026cognitive}. Models often still rely on statistical associations between observable facial, acoustic, or linguistic cues and emotion categories~\citep{jiang2025mme,yeo2025beyond,chen2026audio}, while accurate perception alone does not ensure that perceptual evidence is explicitly organized into cognition-level appraisal reasoning~\citep{ong2015affective,houlihan2023emotion,yeo2025beyond}. Such \textit{\textbf{cognitive misalignment}} becomes particularly problematic when multimodal evidence is ambiguous, conflicting, or context-dependent, and may contribute to inappropriate empathic responses~\citep{wang2022empathetic} or unreliable emotion-informed decisions in mental-health settings~\citep{zhang2023emotion}. Yet existing multimodal emotion datasets largely provide emotion labels or perception-level descriptions~\citep{lian2026merbench,cheng2024emotion,lian2025affectgpt}, while conventional metrics mainly evaluate final predictions, leaving cognition-level reasoning insufficiently supervised and assessed.

\begin{figure}[!t]
    \centering
    \includegraphics[width=0.95\linewidth]{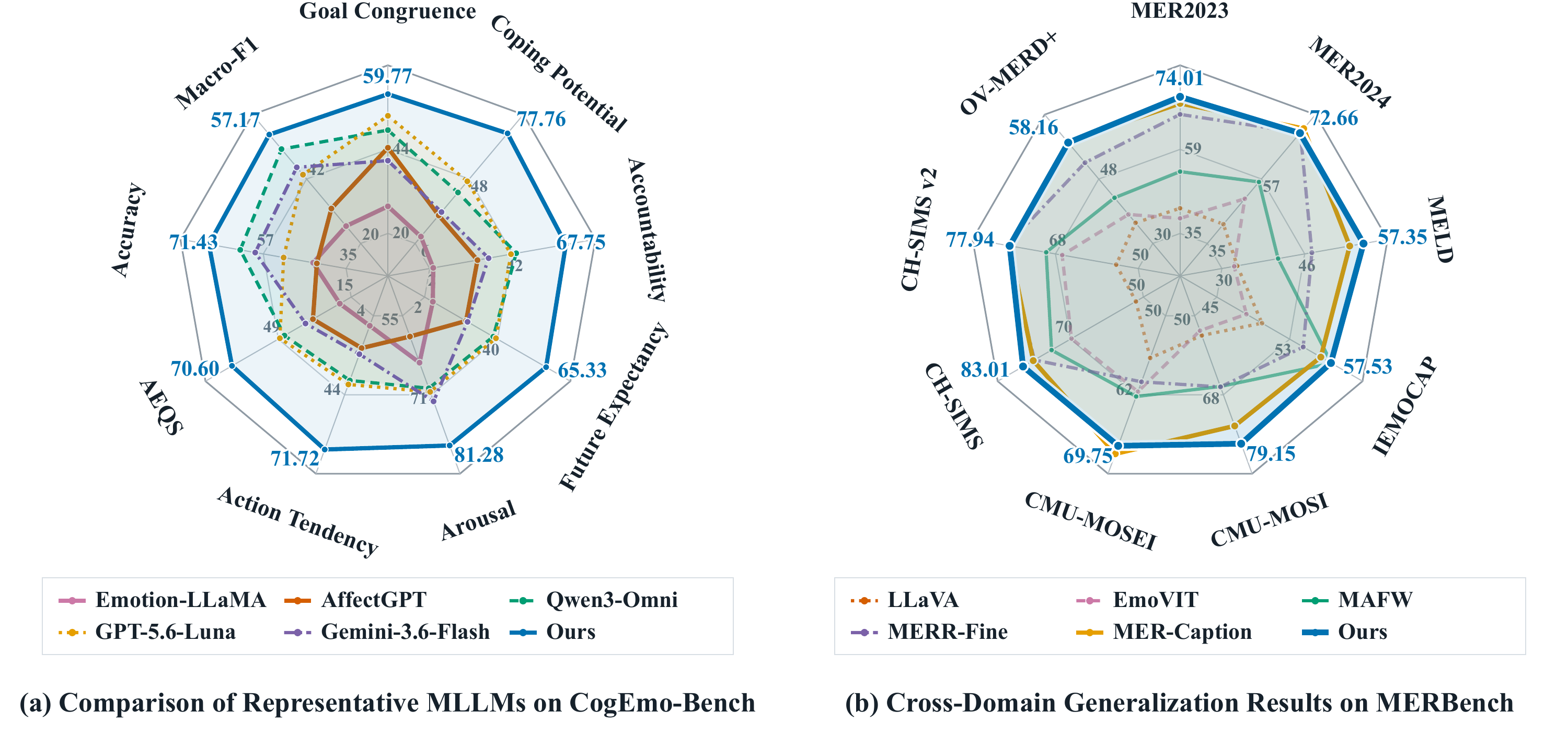} 
    \caption{Performance comparison of representative MLLMs on cognitive appraisal reasoning and cross-domain emotion understanding.}
    \label{fig:motivation}
\end{figure}

From a psychological perspective,  we model cognitive appraisal as a key intermediate process in emotion understanding, rather than treating emotion as a direct cue-to-label mapping. Specifically, we characterize appraisal reasoning through six complementary dimensions: goal congruence, coping potential, accountability, future expectancy, arousal, and action tendency~\citep{roseman1991appraisal,yeo2024associations}, providing an interpretable structure for reasoning about why an emotion arises. However, existing emotion-oriented MLLMs provide limited support for cognition-specific specialization, and scaling model capacity alone is inefficient for targeted cognitive adaptation.

The main contributions of this paper are summarized as follows:

\begin{enumerate}
[leftmargin=*,label=\textcircled{\arabic*}]
\item We introduce a \textbf{perception-to-appraisal paradigm} for multimodal emotion understanding, shifting from direct cue-to-label prediction toward reasoning about how perceptual evidence is cognitively interpreted before emotion arises. To support this paradigm, we construct \textbf{CogEmo-40K}, a large-scale human-aligned instruction-tuning dataset for cognition-grounded emotion reasoning.

\item We propose \textbf{CogEmo-MoE}, a lightweight sparse MLLM for cognitive
appraisal reasoning with perception-preserving adaptation. Instead of scaling up or broadly fine-tuning the backbone, we preserve established multimodal perception and specialize only sparse experts for appraisal-specific adaptation, enabling strong appraisal reasoning within a compact model.

\item  
We develop \textbf{CogEmo-Bench},  a fine-grained benchmark with an Appraisal Evidence Quality Score (AEQS) for evaluating cognitive appraisal reasoning and emotion recognition. Across diverse open- and closed-source MLLMs, CogEmo-MoE improves AEQS over GPT-5.2 from \textbf{\(\mathbf{56.30}\!\rightarrow\!\mathbf{70.60}\)} and accuracy over Qwen3-Omni from \textbf{\(\mathbf{63.10\%}\!\rightarrow\!\mathbf{71.43\%}\)}, revealing both the strengths and limitations of current models in cognitive emotion understanding.
\end{enumerate}

\begin{figure}[!h]
    \centering
    \includegraphics[width=\linewidth]{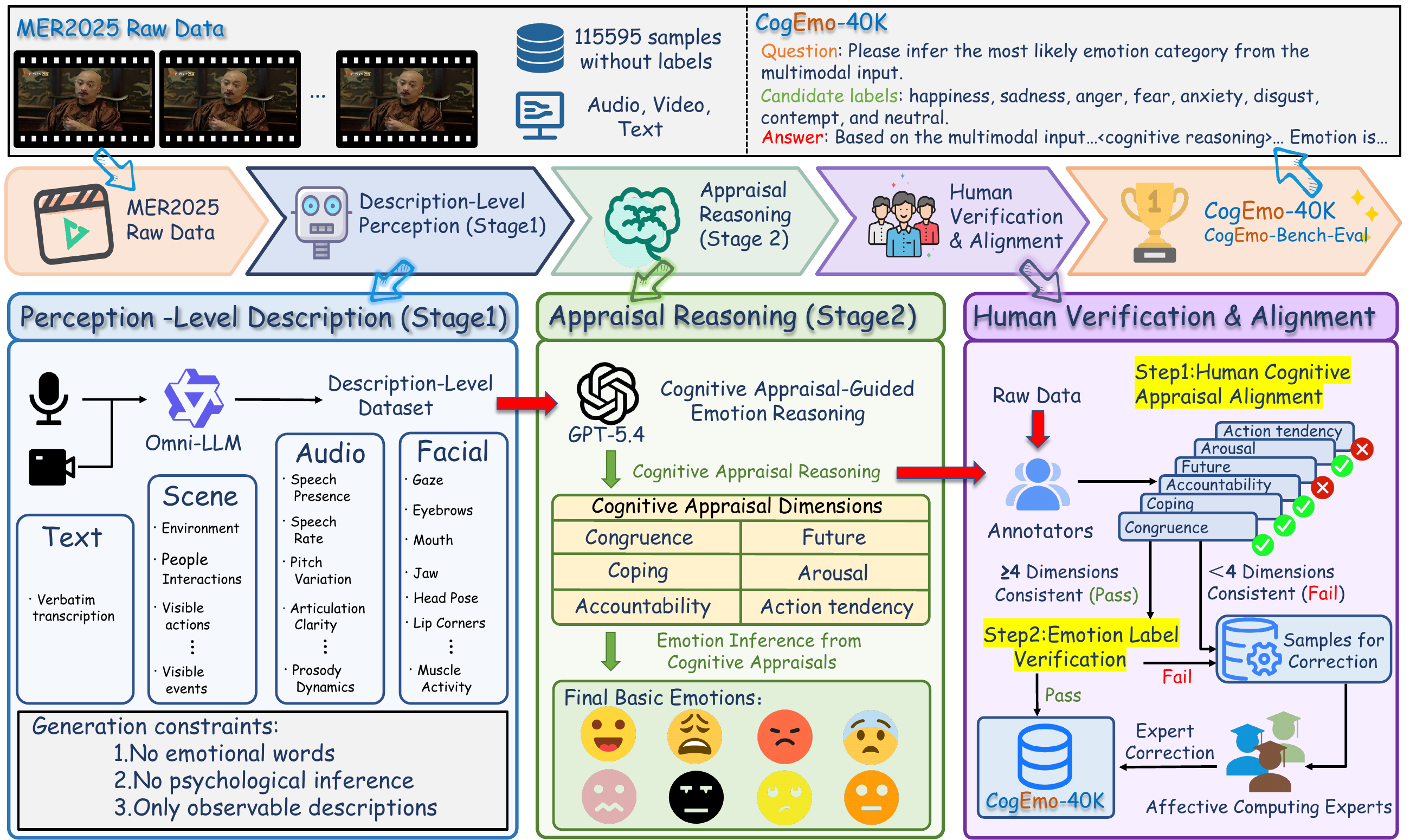}
    \caption{Illustration of CogEmo-40K Dataset Construction.}
    \label{fig:dataset_construction}
\end{figure}

\section{CogEmo-40K: Cognition-Grounded Multimodal Emotion Reasoning Dataset}
\label{sec:dataset}

\subsection{Dataset Overview}
\label{sec:dataset_overview}



Existing multimodal emotion datasets mainly provide emotion labels or perception-level descriptions, offering limited supervision for the cognitive appraisal processes underlying emotional responses. To address this gap, we construct \textbf{CogEmo-40K}, a cognition-grounded multimodal emotion reasoning dataset. Following the screening protocol of AffectGPT~\citep{lian2025affectgpt}, we annotate the same 115{,}595 MER2025~\citep{lian2025mer} source samples used by MERCaption~\citep{lian2025affectgpt} through a two-stage perception-to-appraisal pipeline, followed by human-guided alignment and refinement to obtain approximately 40K high-quality instances. A held-out subset forms \textbf{CogEmo-Bench-Eval}, providing over 6,000 human-verified dimension-level appraisal references across six appraisal dimensions. The overall construction pipeline is illustrated in \styleref{Figure~\ref{fig:dataset_construction}}. Each instance explicitly follows a \textit{multimodal observation $\rightarrow$ cognitive appraisal $\rightarrow$ emotion state} reasoning process, providing supervision not only for \textit{\textbf{what emotion is expressed}}, but also for \textit{\textbf{how emotional responses emerge from interpreted situations}}. Detailed dataset statistics are provided in Appendix~\ref{app:dataset_statistics}.

\subsection{Two-Stage Perception-to-Appraisal Data Generation}
\label{sec:two_stage_generation}
As illustrated in \styleref{Figure~\ref{fig:dataset_construction}}, we decompose annotation generation into \textit{perception} and \textit{appraisal} to reduce the entanglement between perceptual description and cognitive interpretation, while discouraging shortcut cue-label associations.

\textbf{Stage 1: Perception-Level Description.}
The first stage converts multimodal inputs into emotion-agnostic descriptions of observable evidence across four sources: \textit{Scene}, \textit{Face}, \textit{Speaker Audio}, and \textit{Transcript}. Emotion labels and latent psychological interpretations are explicitly excluded to prevent appraisal leakage, yielding an evidence-grounded representation for subsequent cognitive reasoning.

\textbf{Stage 2: Cognition-Level Appraisal Reasoning.}
The second stage interprets the perceptual evidence through six appraisal dimensions motivated by cognitive appraisal theory: \textit{Goal Congruence}, \textit{Coping}, \textit{Accountability}, \textit{Future Expectancy}, \textit{Arousal}, and \textit{Action Tendency}. Rather than directly inferring an emotion label from observable cues, the model is required to explain how the perceived situation is evaluated along these dimensions before deriving the emotional response. Each appraisal is grounded in supporting evidence from perception-level description, yielding an explicit reasoning trajectory from multimodal observation to cognitive interpretation and finally to emotion state.


\subsection{Human-Guided Cognitive Alignment}
\label{sec:quality_control}

Reliable cognition-grounded supervision requires consistency across the full \textit{perception--appraisal--emotion} chain. We therefore conduct human-guided alignment at perceptual, cognitive, and emotional levels.

\textbf{Perception-Level Quality Assessment.}
We randomly sample 1,000 generated descriptions and ask five researchers with backgrounds in affective computing to evaluate their \textit{Accuracy}, \textit{Completeness}, and \textit{Objectivity} on a five-point scale, following \styleref{Appendix~\ref{app:description_guideline}}. The average score for each criterion exceeds 4.0, indicating that the generated descriptions provide reliable and largely emotion-agnostic perceptual evidence.

\textbf{Cognitive Appraisal and Emotion Alignment.}
We recruit 300 trained volunteers to assess whether the generated cognitive appraisal reasoning is supported by the original multimodal evidence and aligned with plausible human interpretations. For each sample, volunteers compare the generated appraisals with human interpretations across the six cognitive appraisal dimensions, while also verifying whether the final emotion label is consistent with both the multimodal context and the inferred cognitive appraisal reasoning. Rather than directly discarding inconsistent, uncertain, or ambiguous cases, we adopt a human-in-the-loop refinement process to revise the cognitive reasoning and emotion labels based on human feedback, followed by further expert review. This process preserves challenging samples while improving the consistency between cognitive appraisals and emotion labels.

\section{CogEmo-MoE: Model Design}
\label{sec:model}

\begin{figure}[t]
\centering
\includegraphics[width=\linewidth]{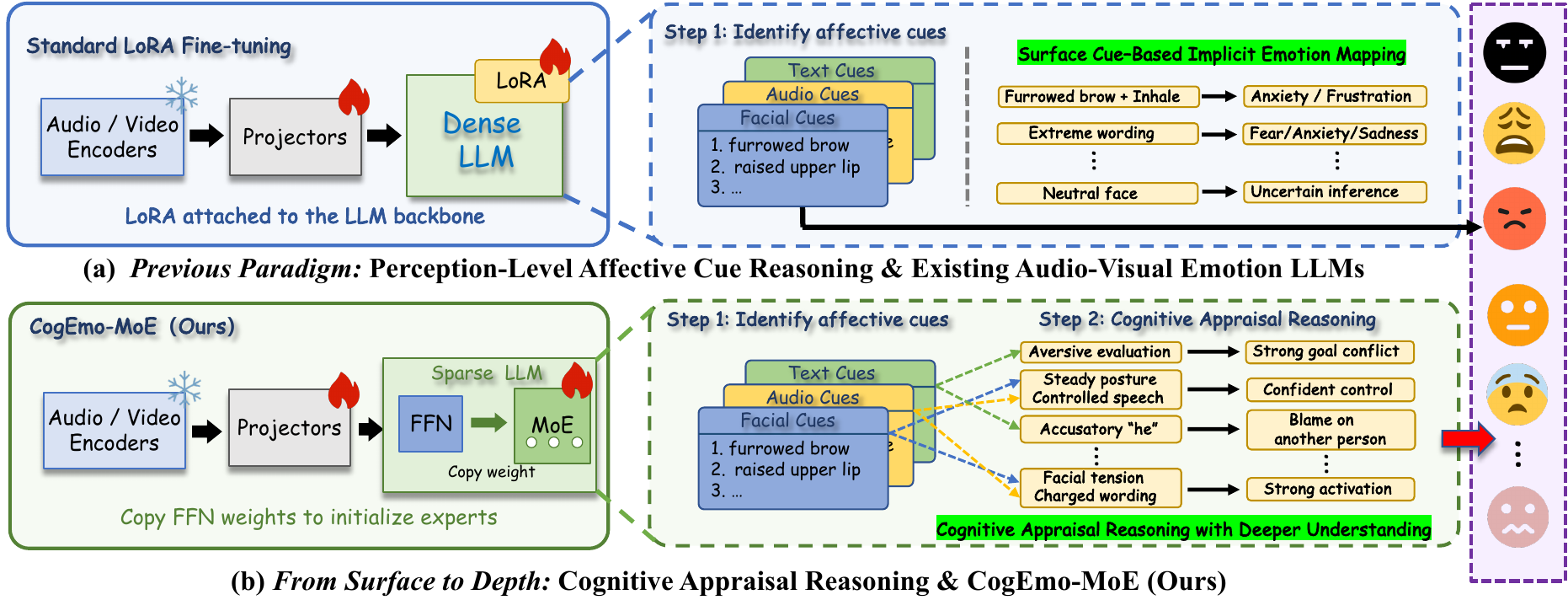}
\caption{Comparison of surface cue-based emotion reasoning and our cognitive appraisal-based CogEmo-MoE framework.}
\label{fig:model}
\end{figure}

\subsection{Architecture Overview}
\label{sec:architecture}
Existing multimodal emotion models often rely on large backbones for affective understanding. We instead explore whether compact models can acquire effective cognitive appraisal through targeted specialization, and introduce \textbf{CogEmo-MoE}, a lightweight sparse model built upon MoE-LLaVA.
It adopts \texttt{Qwen-1.8B} as the language backbone and \texttt{CLIP}/\texttt{CLAP} as the visual/audio encoders. As shown in \styleref{Fig.~\ref{fig:model}}, sparse experts initialized from the corresponding FFN weights are specialized for cognitive appraisal reasoning. Given video $V$, audio $A$, and instruction $I$, multimodal features are projected into the language embedding space and combined with the instruction embeddings:
\vspace{-1pt}
\begin{equation}
Y =
LM_{\mathrm{MoE}}
\left(
[P_v(f_{\mathrm{CLIP}}(V));
P_a(f_{\mathrm{CLAP}}(A));
E(I)]
\right),
\end{equation}
\vspace{-1pt}
where $P_v(\cdot)$ and $P_a(\cdot)$ are modality-specific projectors, $E(I)$ is the instruction embedding, and $LM_{\mathrm{MoE}}$ denotes the MoE-augmented language backbone for cognition-specific reasoning.

\subsection{Sparse Cognitive Expert Adaptation}
\label{sec:moe}
Cognitive appraisal reasoning builds on earlier multimodal perception. We therefore interleave sparse MoE modules across the Transformer backbone. Each MoE layer contains multiple experts and a lightweight router, which computes the selection probability for expert $e$ given hidden state $h$:
\vspace{-1pt}
\begin{equation}
p(e|h)=\mathrm{Softmax}(W_rh),
\end{equation}
\vspace{-1pt}
where $W_r$ denotes the routing parameters. Only the top-$k$ experts are
activated for each input:
\vspace{-1pt}
\begin{equation}
\mathrm{MoE}(h)
=
\sum\nolimits_{i\in \mathrm{TopK}(p)}
p_iE_i(h),
\end{equation}
\vspace{-1pt}
where $E_i(\cdot)$ denotes the $i$-th expert and $p_i$ is its routing weight.


\subsection{Progressive Perception-to-Cognition Training}
\label{sec:training}

CogEmo-MoE follows a progressive three-stage training strategy. All stages are optimized with the same autoregressive objective, while differing in supervision granularity and trainable modules. Given the stage-specific dataset $\mathcal{D}_s$, the training objective is
\vspace{-1pt}
\begin{equation}
\mathcal{L}_s
=
-\sum\nolimits_{(X,I,Y)\in\mathcal{D}_s}
\sum\nolimits_{t=1}^{|Y|}
\log p_{\theta_s}
\left(
y_t \mid y_{<t}, X, I
\right),
\qquad s\in\{1,2,3\},
\label{eq:progressive_training}
\end{equation}
\vspace{-1pt}
where $X$, $I$, and $Y$ denote the multimodal input, instruction, and stage-specific target, respectively, and $\theta_s$ are the trainable parameters. The training data used are summarized in \styleref{Table~\ref{tab:training_data}}. \textbf{MERR-Coarse} is a coarse-grained face--audio emotion dataset from Emotion-LLaMA~\citep{cheng2024emotionllama}, whereas our \textbf{CogEmo-Desc} comprises fine-grained perception-level descriptions generated during the automatic annotation process.


\textbf{Stage 1: Multimodal Alignment.}
Using \textbf{MERR-Coarse}, we optimize only the modality-specific projectors, while keeping the pretrained encoders and language backbone frozen, to align visual and audio representations with the language embedding space.



\textbf{Stage 2: Perception Enhancement.}
Starting from the aligned model, we introduce \textbf{CogEmo-Desc} as fine-grained description-level supervision to strengthen the model's perception of facial and acoustic details. During this stage, both the language backbone and modality-specific projectors are unfrozen and jointly optimized.

\noindent
\begin{minipage}[c]{0.48\linewidth}


\textbf{Stage 3: Cognitive Reasoning Adaptation.}
Finally, we use \textbf{CogEmo-40K} as cognition-level supervision to specialize the sparse experts for appraisal-based emotion reasoning on top of the multimodal perception acquired in the previous stages. During this stage, the language backbone is frozen and only the sparse expert modules are optimized, concentrating adaptation on cognition-specific reasoning.

\end{minipage}
\hfill
\begin{minipage}[c]{0.45\linewidth}
\centering

\captionof{table}{Training data used in the three stages of CogEmo-MoE.}
\label{tab:training_data}

\vspace{-2pt}

\setlength{\tabcolsep}{6pt}
\renewcommand{\arraystretch}{1.15}
\renewcommand{\arraystretch}{1.05}

\begin{tabular*}{\linewidth}{
@{\hspace{5pt}}
l
@{\extracolsep{\fill}}
l
r
@{\hspace{5pt}}
}
\toprule
\textbf{Usage}
& \textbf{Source}
& \textbf{\#Samples}
\\
\midrule
Stage I   & MERR-Coarse & 28,618 \\
Stage II  & CogEmo-Desc & 114,587 \\
Stage III & CogEmo-40K  & 38,532 \\
\bottomrule
\end{tabular*}

\end{minipage}

\vspace{4pt}

\section{CogEmo-Bench: Cognition-Grounded Emotion Understanding Benchmark}
\label{sec:benchmark}

\subsection{CogEmo-Bench Overview}
The introduced \textbf{CogEmo-Bench} jointly evaluates cognitive appraisal reasoning and conventional emotion recognition. An \textit{\textbf{Appraisal Extractor}} normalizes free-form MLLM responses into the six appraisal dimensions defined in Sec.~\ref{sec:two_stage_generation}, without judging correctness or adding evidence. The extracted appraisals are compared with human-verified references using the \textbf{Appraisal Evidence Quality Score (AEQS)}, while final emotion predictions are evaluated with standard classification metrics.



\subsection{Evaluation Metrics}
\label{sec:benchmark_metrics}

Conventional emotion metrics evaluate final predictions but do not reveal whether the underlying cognitive evidence is well aligned with human interpretation. We therefore introduce a new metric to evaluate appraisal evidence across the six cognitive dimensions.

\textbf{Appraisal Evidence Quality Score (AEQS).} For sample $i$ and appraisal dimension $d$, let $\hat{e}_{i,d}$ and $e_{i,d}$ denote the extracted model evidence and human-verified reference, respectively. Using a fixed embedding encoder $\phi(\cdot)$, the evidence similarity is defined as
\vspace{-1pt}
\begin{equation}
s_{i,d}
=
\cos\!\left(
\phi(\hat{e}_{i,d}),
\phi(e_{i,d})
\right).
\label{eq:appraisal_similarity}
\end{equation}
\vspace{-1pt}
If only one of $\hat{e}_{i,d}$ and $e_{i,d}$ is empty, we assign $s_{i,d}=0$; if both are empty, the sample is excluded from the corresponding dimension. Let $\mathcal{S}_d$ denote the valid sample set for dimension $d$. The dimension-level score is
\vspace{-1pt}
\begin{equation}
S_d
=
\frac{1}{|\mathcal{S}_d|}
\sum\nolimits_{i\in\mathcal{S}_d} s_{i,d}.
\label{eq:dimension_score}
\end{equation}
\vspace{-1pt}
AEQS is then computed by averaging across the six appraisal dimensions:
\vspace{-1pt}
\begin{equation}
\mathrm{AEQS}
=
\frac{\alpha}{D}
\sum\nolimits_{d=1}^{D} S_d,
\label{eq:aeqs}
\end{equation}
\vspace{-1pt}
where $D$ denotes the number of appraisal dimensions and $\alpha$ is a scaling
factor; we set $D=6$ and $\alpha=100$ in this work. Higher AEQS indicates stronger semantic alignment between model-generated appraisal evidence and human-verified cognitive references.

\textbf{Emotion Recognition.}
We additionally report standard \textbf{Accuracy} and \textbf{Macro-F1} for final emotion prediction, with detailed computation provided in Appendix~\ref{app:bench_metrics}.

\section{Results and Observations}

\paragraph{Main results.}  Detailed implementation information can be found in \styleref{Appendix~\ref{app:implementation_details}}.
\styleref{Table~\ref{tab:cogemo_bench}} compares representative MLLMs on both cognitive appraisal reasoning and emotion recognition. 
The results show substantial differences across models in AEQS and final emotion performance. 
Some models achieve competitive emotion classification despite relatively weak appraisal reasoning, while larger general-purpose and proprietary models generally perform better on both aspects. 
CogEmo-MoE achieves a favorable balance between appraisal reasoning and emotion recognition despite its compact model scale.
Specifically, it improves AEQS over GPT-5.2 from \textbf{\(\mathbf{56.30}\!\rightarrow\!\mathbf{70.60}\)}, while outperforming Qwen3-Omni in accuracy from \textbf{\(\mathbf{63.10\%}\!\rightarrow\!\mathbf{71.43\%}\)} and Macro-F1 from \textbf{\(\mathbf{52.19\%}\!\rightarrow\!\mathbf{57.17\%}\)}. Additional model-size comparisons are provided in \styleref{Appendix~\ref{app:model_scale}}.


\noindent\textbf{Obs. 1: Emotion recognition does not necessarily imply cognitive understanding.}
VideoChat-Flash~\citep{li2025videochat} achieves strong emotion prediction despite limited explicit appraisal reasoning, while emotion-specialized models such as AffectGPT~\citep{lian2025affectgpt} and Emotion-LLaMA~\citep{cheng2024emotionllama} also show limited transfer to cognition-level understanding. These results indicate that emotion recognition and surface cue-to-label mapping cannot substitute for cognitive appraisal reasoning about why emotions arise. By contrast, cognition-oriented supervision enables our CogEmo-MoE to improve both appraisal reasoning and emotion discrimination within a compact model.

\noindent\textbf{Obs. 2: True emotion intelligence requires adaptive reasoning across different emotion spaces.} Strong performance within a fixed label space does not necessarily indicate robust emotion understanding. A capable emotion model should adapt its judgments to task-specific emotion taxonomies rather than rely on surface-level cue-to-label mappings. Although open-vocabulary approaches such as AffectGPT~\citep{lian2025affectgpt} improve output flexibility, evaluation protocols that count a prediction as correct whenever the target emotion appears among multiple generated candidates may overestimate performance without establishing that the model selected the most appropriate emotion. In contrast, as shown in \styleref{Fig.~\ref{fig:motivation}(b)} and detailed in Appendix~\ref{app:cross_domain}, our results show that cognition-oriented reasoning transfers across different emotion spaces and consistently improves emotion recognition tasks, suggesting that cognitive appraisal provides a more generalizable basis for emotion understanding.

\begin{table*}[t]
\centering
\caption{
Comparison of different MLLMs on \textbf{CogEmo-Bench}.
A, V, and T denote audio, visual, and textual modalities, respectively. Cong., Cop., Account., Fut., Aro., and Act. denote Goal congruence, Coping potential, Accountability, Future expectancy, Arousal, and Action tendency.}
\label{tab:cogemo_bench}

\vspace{3.5pt}
\setlength{\tabcolsep}{4.5pt}
\renewcommand{\arraystretch}{1.10}

\resizebox{\textwidth}{!}{%
\begin{tabular}{lccc|ccccccc|cc}
\toprule

\textbf{Model}
& \textbf{A}
& \textbf{V}
& \textbf{T}
& \textbf{Cong.}
& \textbf{Cop.}
& \textbf{Account.}
& \textbf{Fut.}
& \textbf{Aro.}
& \textbf{Act.}
& \textbf{AEQS}
& \textbf{Acc.}
& \textbf{M-F1}
\\

\midrule


\rowcolor{groupteal}
\multicolumn{13}{c}{\textit{Open-source MLLMs}}
\\

OneLLM~\citep{han2024onellm}
& $\checkmark$ & $\times$ & $\checkmark$
& 31.47 & 8.02 & 10.10 & 9.59 & 30.90 & 6.81
& 16.15 & 30.16 & 26.03
\\

SECap~\citep{xu2024secap}
& $\checkmark$ & $\times$ & $\checkmark$
& 11.12 & 7.42 & 4.59 & 3.97 & 42.26 & 4.23
& 12.26 & 30.85 & 21.25
\\

Qwen2-Audio~\citep{chu2024qwen2audio}
& $\checkmark$ & $\times$ & $\checkmark$
& 39.27 & 18.16 & 22.26 & 24.85 & 30.55 & 15.60
& 25.12 & 43.55 & 32.68
\\

SALMONN~\citep{tang2024salmonn}
& $\checkmark$ & $\times$ & $\checkmark$
& 42.87 & 18.20 & 20.18 & 24.10 & 27.17 & 10.67
& 23.87 & 43.69 & 33.43
\\

Otter~\citep{li2025otter}
& $\times$ & $\checkmark$ & $\checkmark$
& 26.09 & 3.83 & 8.90 & 3.89 & 15.65 & 3.58
& 10.32 & 36.61 & 24.69
\\

LLaMA-VID~\citep{li2024llama}
& $\times$ & $\checkmark$ & $\checkmark$
& 47.73 & 16.35 & 20.33 & 21.62 & 39.77 & 13.79
& 26.60 & 36.71 & 31.84
\\

VideoChat-Flash~\citep{li2025videochat}
& $\times$ & $\checkmark$ & $\checkmark$
& 6.82 & 1.22 & 1.26 & 1.75 & 4.32 & 0.78
& 2.69 & 55.36 & 39.11
\\

Chat-UniVi~\citep{jin2024chat}
& $\times$ & $\checkmark$ & $\checkmark$
& 49.96 & 22.80 & 26.66 & 27.92 & 46.63 & 19.73
& 32.28 & 34.03 & 29.86
\\

mPLUG-Owl3~\citep{ye2025mplug}
& $\times$ & $\checkmark$ & $\checkmark$
& 47.78 & 18.53 & 23.11 & 27.49 & 58.21 & 17.30
& 32.07 & 45.54 & 34.72
\\

Qwen3.8-Max~\citep{team2026qwen3}
& $\times$ & $\checkmark$ & $\checkmark$
& 51.30 & 44.37 & 47.12 & 42.56 & 67.71 & 43.94
& 49.50 & 57.04 & 46.29
\\

Kimi-K3~\citep{kimi2026k3}
& $\times$ & $\checkmark$ & $\checkmark$
& 52.02 & 43.58 & 45.23 & 40.54 & 68.38 & 42.42
& 48.70 & 46.13 & 39.04
\\

PandaGPT~\citep{su2023pandagpt}
& $\checkmark$ & $\checkmark$ & $\checkmark$
& 50.97 & 14.65 & 19.92 & 21.22 & 32.92 & 13.55
& 25.54 & 25.79 & 18.78
\\

Emotion-LLaMA~\citep{cheng2024emotionllama}
& $\checkmark$ & $\checkmark$ & $\checkmark$
& 27.74 & 10.42 & 3.71 & 6.24 & 64.49 & 9.04
& 20.27 & 43.75 & 25.90
\\

AffectGPT~\citep{lian2025affectgpt}
& $\checkmark$ & $\checkmark$ & $\checkmark$
& 44.46 & 24.26 & 25.05 & 23.58 & 59.20 & 20.39
& 32.83 & 42.76 & 31.90
\\

Qwen3-Omni~\citep{xu2025qwen3omni}
& $\checkmark$ & $\checkmark$ & $\checkmark$
& 49.50 & 39.21 & 43.74 & 37.79 & 69.69 & 36.76
& 46.12 & 63.10 & 52.19
\\

\midrule


\rowcolor{groupteal}
\multicolumn{13}{c}{\textit{Closed-source MLLMs}}
\\

GPT-5.2~\citep{openai2025gpt52}
& $\times$ & $\checkmark$ & $\checkmark$
& 55.89 & 56.78 & 53.36 & 48.37 & 70.34 & 53.07
& 56.30 & 46.33 & 36.92
\\

GPT-5.6-Luna~\citep{openai2026gpt56}
& $\times$ & $\checkmark$ & $\checkmark$
& 53.57 & 46.48 & 41.23 & 39.18 & 70.41 & 38.76
& 48.27 & 51.59 & 43.46
\\

Grok-4.3~\citep{xai2026grok43}
& $\times$ & $\checkmark$ & $\checkmark$
& 51.67 & 36.80 & 40.67 & 37.82 & 69.46 & 36.88
& 45.55 & 46.83 & 39.79
\\

Gemini-3.6-Flash~\citep{google2026gemini36flash}
& $\checkmark$ & $\checkmark$ & $\checkmark$
& 40.79 & 26.42 & 30.44 & 24.36 & 72.33 & 23.45
& 36.30 & 59.13 & 45.99
\\

\midrule


\rowcolor{groupteal}
\multicolumn{13}{c}{\textit{Our CogEmo Models}}
\\

CogEmo-MoE
& $\checkmark$ & $\times$ & $\checkmark$
& 58.68
& 77.04
& 67.44
& 65.04
& 81.03
& \textbf{71.81}
& 70.17
& 63.59
& 50.18
\\

CogEmo-MoE
& $\times$ & $\checkmark$ & $\checkmark$
& 59.56
& 77.56
& 66.88
& 64.59
& 80.32
& 71.26
& 70.03
& 62.70
& 50.07
\\

CogEmo-MoE
& $\checkmark$ & $\checkmark$ & $\checkmark$
& \textbf{59.77}
& \textbf{77.76}
& \textbf{67.75}
& \textbf{65.33}
& \textbf{81.28}
& 71.72
& \textbf{70.60}
& \textbf{71.43}
& \textbf{57.17}
\\

\bottomrule
\end{tabular}%
}
\vspace{2pt}
\end{table*}

\begin{figure}[htbp]
    \centering
    \includegraphics[width=\linewidth]{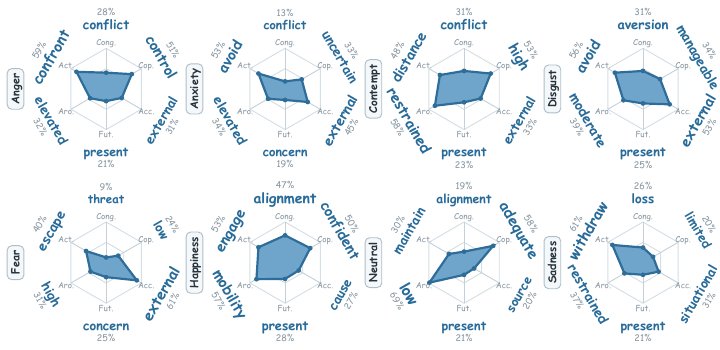}
    \caption{
   Emotion-Specific Cognitive Language Profiles in CogEmo-40K. 
    Radar plots show six-dimensional appraisal patterns, with radial extent and label size indicating phrase prevalence.
    }
    \label{fig:cognitive-profiles}
\end{figure}

\paragraph{Cognitive Language Profiles.}
We further analyze the structured outputs of the fixed Appraisal Extractor by aggregating recurring phrase-level patterns for each emotion and appraisal dimension. \styleref{Figure~\ref{fig:cognitive-profiles}} visualizes the resulting emotion-specific profiles across six cognitive dimensions, where radial extent and label size indicate the prevalence of the most frequent phrase within each emotion category. The surrounding terms summarize the corresponding appraisal semantics. These percentages reflect phrase prevalence rather than appraisal intensity. The original top-3 phrases and their occurrence rates are reported in Appendix~\ref{app:cognitive_profiles}.

\noindent\textbf{Obs. 3: Emotion categories emerge from structured configurations of cognitive appraisals rather than isolated affective cues.} As shown in \styleref{Figure~\ref{fig:cognitive-profiles}}, emotions exhibit distinct joint appraisal profiles even when individual dimensions overlap. For example, anger is characterized by conflict, maintained control, external attribution, elevated arousal, and confrontation, whereas sadness is associated with loss, limited coping, situational attribution, restrained arousal, and withdrawal. Likewise, fear emphasizes threat, low coping, high arousal, and escape, while anxiety is more associated with conflict, uncertain coping, elevated arousal, and avoidance. These patterns suggest that emotion differentiation is better characterized by configurations of cognitive appraisals than by isolated affective cues.

\begin{figure*}[htbp]
    \centering
    \includegraphics[width=\textwidth]{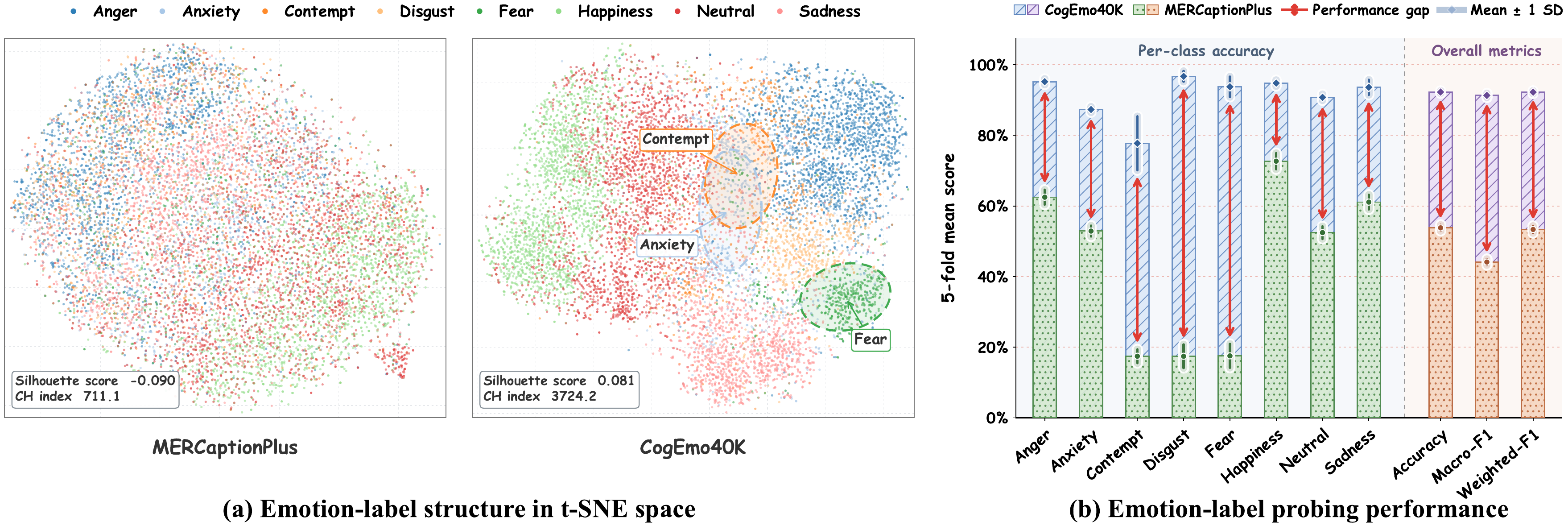}
    \caption{
    Emotion Alignment of Cognition-Grounded Supervision.
    We compare CogEmo-40K and MERCaption+ on their intersected samples under the same eight-way emotion partition.}
    \label{fig:emotion_separability}
\end{figure*}

\paragraph{Emotion Alignment of Cognition-Grounded Supervision.}
On the samples shared by CogEmo-40K and MERCaption+~\citep{lian2025affectgpt}, we compare how the two forms of supervision encode emotion-related information. As shown in \styleref{Figure~\ref{fig:emotion_separability}(a)}, cognition-grounded reasoning yields clearer emotion-label structure than open-ended descriptions. For \styleref{Figure~\ref{fig:emotion_separability}(b)}, we conduct five-fold cross-validation with the same \texttt{XLM-RoBERTa} classification framework for each supervision source. After removing explicit emotion-conclusion sentences, the classifier predicts eight emotion categories from each source's text. CogEmo-40K achieves stronger per-class and overall performance, suggesting that its appraisal reasoning encodes information more closely aligned with the final emotion labels. Detailed settings are provided in Appendix~\ref{app:5fold}.


\noindent
\begin{minipage}[t]{0.48\textwidth}
\vspace{-2pt}

\textbf{Effect of MoE Placement.}
We investigate where the sparse expert layers should be introduced.
We compare four placement strategies, including First-Half, Second-Half,
Interval, and All. The Interval strategy achieves the best overall performance,
raising accuracy from \textbf{\(\mathbf{64.29\%}\!\rightarrow\!\mathbf{71.43\%}\)}
over First-Half placement and supporting our design of inserting MoE layers at
alternating Transformer blocks for more balanced cognition-specific adaptation.

\vspace{3pt}

\textbf{Dense vs. Sparse Experts.}
We further compare the dense baseline with different sparse routing strategies.
Simply increasing the number of experts with Top-1 routing does not improve
performance. In contrast, Top-2 routing improves both emotion recognition and
cognitive appraisal quality, including accuracy from \textbf{\(\mathbf{65.97\%}\!\rightarrow\!\mathbf{71.43\%}\)} and Macro-F1 from \textbf{\(\mathbf{52.28\%}\!\rightarrow\!\mathbf{57.17\%}\)} under the four-expert setting, suggesting that collaborative expert activation is more effective for cognition-level emotion reasoning.

\vspace{3pt}


\end{minipage}
\hfill
\begin{minipage}[t]{0.5\textwidth}
\vspace{-2pt}
\centering


\captionof{table}{
Effect of different MoE placement strategies on \textbf{CogEmo-Bench-Eval}.
}
\label{tab:moe_placement}

\vspace{-6pt}
\renewcommand{\arraystretch}{1.05}

\begin{tabular*}{0.9\linewidth}{
@{\hspace{5pt}}
l
@{\extracolsep{\fill}}
ccc
@{\hspace{5pt}}
}
\toprule
\textbf{Architecture}
& \textbf{AEQS}
& \textbf{Acc.}
& \textbf{M-F1}
\\
\midrule
First-Half  & 68.76 & 64.29 & 50.80 \\
Second-Half & 70.10 & 68.45 & 54.57 \\
Interval    & \textbf{70.60} & \textbf{71.43} & \textbf{57.17} \\
All         & 69.76 & 70.04 & 55.49 \\
\bottomrule
\end{tabular*}

\vspace{2pt}


\captionof{table}{
Comparison of dense and sparse expert configurations on \textbf{CogEmo-Bench-Eval}.
}
\label{tab:expert_ablation}

\vspace{2pt}
\renewcommand{\arraystretch}{1.06}

\begin{tabular*}{0.9\linewidth}{
@{\hspace{5pt}}
l
@{\extracolsep{\fill}}
ccc
@{\hspace{5pt}}
}
\toprule

\textbf{Configuration}
& \textbf{AEQS}
& \textbf{Acc.}
& \textbf{M-F1}
\\

\midrule

\rowcolor{groupteal}
\multicolumn{4}{c}{\textit{Dense Model}}
\\

Baseline
& 70.48
& 67.26
& 53.27
\\

\midrule

\rowcolor{groupteal}
\multicolumn{4}{c}{\textit{Sparse MoE}}
\\

4E / Top-1
& 70.02
& 65.97
& 52.28
\\

4E / Top-2
& \textbf{70.60}
& \textbf{71.43}
& \textbf{57.17}
\\

8E / Top-1
& 69.80
& 64.68
& 51.40
\\

8E / Top-2
& 69.76
& 70.04
& 55.49
\\

\bottomrule
\end{tabular*}

\end{minipage}

\section{Related Work}
\subsection{From Multimodal Perception to Emotion Reasoning}

MLLMs align visual, acoustic, and temporal evidence with language, providing a foundation for open-ended multimodal interpretation~\citep{han2024onellm,li2024llama,chu2024qwen2audio}. In affective computing, earlier systems mapped fused multimodal representations to predefined emotion categories~\citep{zhang2024deep,zadeh-etal-2017-tensor,tsai-etal-2019-multimodal}, while contextual models demonstrated the importance of scenes and conversational history beyond isolated expressions~\citep{kosti2017emotion,lee2019context,poria2017context}. Emotion-oriented MLLMs further generate explanations alongside predictions~\citep{cheng2024emotionllama,lian2025affectgpt,zhao2025r1}, supported by explanation-rich datasets and unified benchmarks~\citep{cheng2024emotionllama,lian2025affectgpt,lian2026merbench,zhang2026mme}. Nevertheless, current supervision and evaluation still emphasize final labels, observable cues, or textual plausibility. Because fluent rationales can be unstable and benchmark formats can admit shortcuts~\citep{jiang2025mme,salido2025none}, explanation generation alone does not establish that a model captures how an emotion arises.

\subsection{Cognitive Appraisal as an Intermediate Representation}

Appraisal theories locate emotion not in observable cues alone, but in how an individual evaluates an event with respect to goals, coping potential, responsibility, expected outcomes, action readiness and so on~\citep{roseman1991appraisal,frijda1987emotion,ellsworth2003appraisal,moors2013appraisal}. Empirical and computational studies further link appraisal patterns and abstract event representations to emotion judgments~\citep{yeo2024associations,ong2015affective,skerry2015neural,houlihan2023emotion}. Although appraisal has informed text-based emotion modeling and Theory-of-Mind evaluation~\citep{hofmann2020appraisal,troiano2023dimensional,yeo2025beyond}, it has yet to be integrated systematically with multimodal perception and free-form reasoning. We operationalize appraisal as an explicit intermediate representation between observation and emotion: CogEmo-40K provides perception-to-appraisal supervision, CogEmo-MoE models this transition, and CogEmo-Bench evaluates it.

\section{Conclusion}

We introduce a perception-to-appraisal paradigm for multimodal emotion understanding. CogEmo-40K supplies cognition-grounded supervision, CogEmo-MoE applies sparse appraisal-specific adaptation, and CogEmo-Bench uses AEQS to evaluate appraisal evidence alongside emotion recognition. On CogEmo-Bench, CogEmo-MoE achieves 70.60 AEQS, 71.43\% accuracy, and 57.17\% Macro-F1. Analyses of appraisal language reveal distinct patterns across emotion categories, while cross-domain experiments suggest that cognition-grounded supervision can transfer beyond the training domain. These results support appraisal reasoning as an intermediate representation beyond surface-level cue-to-label emotion recognition.


\subsubsection*{Acknowledgments}
We gratefully acknowledge the psychology researchers and domain experts who provided valuable consultation on cognitive appraisal theory, annotation design, and human cognitive alignment throughout this project. We also thank the affective computing researchers, annotators, and volunteers who contributed to the verification and refinement of CogEmo-40K.

\subsection*{AI use statement}
Generative AI tools were used to improve the clarity, grammar, and readability of the manuscript. All AI-assisted edits were reviewed and revised by the authors, who take full responsibility for the final content.

\subsection*{Reproducibility Statement}
We provide detailed descriptions of the data construction, model training, and evaluation procedures to facilitate reproducibility. The complete two-stage data-generation prompts and human-verification protocols for CogEmo-40K are provided in Appendix~\ref{app:cogemo_details} and Appendix~\ref{app:human_verification}. Model architecture, three-stage training configurations, optimization settings, and training prompts for CogEmo-MoE are reported in Appendix~\ref{app:implementation_details}. For CogEmo-Bench, we provide the evaluation prompts, Appraisal Extractor, metric computation procedure, and model-specific evaluation settings in Appendix~\ref{app:cogemo_bench}. Additional experimental details, including the 5-fold cognition--emotion alignment analysis and cross-domain MERBench evaluation, are provided in Appendix~\ref{app:5fold} and Appendix~\ref{app:cross_domain}, respectively. These details specify the data
sources, model configurations, training hyperparameters, prompts, and evaluation protocols used throughout the experiments.

\subsection*{Ethics Statement}
CogEmo-40K is constructed from the existing MER2025 multimodal dataset rather than from newly collected audiovisual recordings. Human involvement in our study is limited to annotation, verification, and refinement of automatically generated perceptual descriptions, cognitive appraisal reasoning, and emotion labels. Because emotion and cognitive appraisal are inherently context-,
culture-, and individual-dependent, the resulting annotations should be viewed as evidence-grounded interpretations rather than definitive descriptions of an individual's internal psychological state. We therefore caution against using model-generated appraisal reasoning for clinical diagnosis, psychological profiling, or other high-stakes decisions without appropriate human oversight. Potential biases inherited from the source data, including linguistic, cultural, and domain biases, are further discussed in the limitations section.

\bibliography{iclr2025_conference}
\bibliographystyle{iclr2027_conference}


\clearpage
\appendix
\raggedbottom

\section*{APPENDIX CONTENTS}

\setlength{\parindent}{0pt}
\setlength{\parskip}{0pt}

\newcommand{\appcontentsline}[3]{%
  \noindent
  \hyperref[#1]{\textbf{Appendix #2.}}%
  \ #3%
  \leaders\hbox{\kern0.18em.\kern0.18em}\hfill
  \pageref{#1}\par
}

\appcontentsline{app:limitations}{A}{Limitations and Social Impact}
\appcontentsline{app:cogemo_details}{B}{Details of CogEmo-40K Dataset}
\appcontentsline{app:implementation_details}{C}{Implementation Details}
\appcontentsline{app:cogemo_bench}{D}{Details of CogEmo-Bench}
\appcontentsline{app:additional_results}{E}{Additional Experimental Results}

\vspace{1.5em}

\section{LIMITATIONS AND SOCIAL IMPACT}
\label{app:limitations}

While our work advances multimodal emotion understanding from surface-level affective cue recognition toward cognition-grounded appraisal reasoning, several directions remain open for future exploration:

\begin{itemize}

\item First, CogEmo-40K adopts six complementary appraisal dimensions inspired by established cognitive appraisal theories. These dimensions capture important aspects of emotion formation but are not intended to exhaust the full space of cognitive emotion processes. Future work may explore richer appraisal dimensions and alternative theoretical formulations.

\item Second, CogEmo-40K is constructed from existing multimodal video data and therefore remains limited by its current domain, language, cultural, and emotion-space coverage. Extending cognition-grounded emotion understanding to more diverse multilingual, cross-cultural, and open-world settings is an important future direction.

\item Third, CogEmo-Bench relies on a fixed Appraisal Extractor and semantic embedding model for normalizing and comparing free-form reasoning outputs. Although the same evaluation protocol is applied consistently to all models, future work may further investigate evaluator-independent cognition-level metrics.

\end{itemize}

Overall, cognition-grounded emotion understanding offers a promising direction for building more reliable and generalizable affective systems. However, model-generated appraisal interpretations should be treated as evidence-grounded reasoning rather than definitive assessments of an individual's internal psychological state, especially in high-stakes applications. We hope this work encourages further research toward more robust, flexible, and human-aligned emotional intelligence in MLLMs.


\clearpage
\section{Details of CogEmo-40K Dataset}
\label{app:cogemo_details}

\subsection{Source Dataset: MER2025}
The raw data for CogEmo-40K are derived from MER2025~\citep{lian2025mer}, the third edition of the MER challenge, which focuses on integrating large language models with multimodal emotion understanding. MER2025 includes a large-scale collection of multimodal samples without emotion annotations, providing video, audio, facial information, and subtitle text. Following the data screening protocol of AffectGPT~\citep{lian2025affectgpt}, we adopt the same MER2025 source samples used to construct MERCaption~\citep{lian2025affectgpt} as the raw data for CogEmo-40K. Importantly, we do not use the automatically generated descriptions provided by MERCaption. Instead, we treat these source samples as unlabeled multimodal inputs and construct new cognition-grounded reasoning annotations through the two-stage perception-to-appraisal pipeline described below. The generated annotations are further refined through multi-round filtering and human-guided alignment (\styleref{Section~\ref{app:human_verification}}), forming the final CogEmo-40K dataset.

\subsection{Two-Stage Data Generation Prompts}
We adopt a two-stage prompt pipeline to generate CogEmo-40K: Stage~1 produces structured, emotion-free perceptual descriptions from the multimodal inputs; Stage~2 takes these perceptual descriptions as input and performs cognitive appraisal reasoning to derive cognition-grounded emotion interpretations. Stage~1 uses \texttt{Qwen3-Omni-30B-A3B-Instruct}~\citep{xu2025qwen3omni}, while Stage~2 uses \texttt{GPT-5.4}~\citep{openai2026gpt54}. The complete prompts for both stages of data generation are
provided below.

\begin{promptboxdata}{Stage 1: Perception Generation}
{\small
\textbf{[Instruction]}\\
You are given a multimodal input (video + audio).
Your task is to describe the scene, facial behavior, and audio signals in a structured and strictly objective way.

\textbf{[Goal]}
\begin{enumerate}\setlength{\itemsep}{0pt}
\item Provide a brief description of the overall scene.
\item Provide a HIGH-DENSITY, structured description of facial movements.
\item Provide a HIGH-DENSITY, structured description of audio features.
\end{enumerate}

\textbf{[Strict Constraints]}
\begin{itemize}\setlength{\itemsep}{0pt}\setlength{\parskip}{0pt}
\item DO NOT use any emotion words (e.g., happy, sad, angry, fear).
\item Only describe observable visual or acoustic information.
\end{itemize}

\textbf{[Scene Description Rules]}
\begin{itemize}\setlength{\itemsep}{0pt}\setlength{\parskip}{0pt}
\item Describe environment, people, and interactions
\item Include visible actions only
\item Keep it brief (1--2 sentences)
\item Use neutral and factual language
\end{itemize}

\textbf{[Facial Description Rules]}\\
You MUST describe ALL facial feature categories below.
For each category include: (1) state; (2) degree (slight / moderate / strong / stable); (3) dynamics (no change / gradual change / fluctuating).

Required categories:
\begin{itemize}\setlength{\itemsep}{0pt}\setlength{\parskip}{0pt}
\item Gaze: direction, stability, shifts
\item Eyes: openness, eyelid tension, blinking pattern, asymmetry
\item Eyebrows: position, movement, contraction or relaxation, symmetry
\item Mouth: opening degree, lip tension, corner direction, lip compression
\item Jaw: tension level, clenching, movement
\item Facial muscles: overall tension, rigidity, relaxation
\item Left--Right asymmetry: differences in facial movement or tension
\item Head pose: orientation, tilt, movement
\item Temporal dynamics: changes over time (start $\rightarrow$ middle $\rightarrow$ end)
\item Overall facial mobility: level of movement intensity (high / medium / low)
\end{itemize}

You MUST additionally describe fine-grained facial action components using observable anatomical or geometric descriptors, including but not limited to:
\begin{itemize}\setlength{\itemsep}{0pt}\setlength{\parskip}{0pt}
\item Lip corners: upturned / downturned / stretched / retracted
\item Cheek region: raised / tightened / relaxed / bulging
\item Eye aperture: widened / narrowed / squinting / partially closed
\item Eyelid behavior: lifted / lowered / tightened / compressed
\item Nasolabial region: deepened / relaxed
\item Teeth visibility: visible / partially visible / not visible
\item Orbital tension: increased / reduced / stable
\item Facial symmetry changes during motion: symmetric / asymmetric activation
\end{itemize}
Use only visible structural descriptions; do NOT label emotions or interpret meaning.
These features must be embedded into Mouth / Eyes / Facial muscles sections.

\textbf{[Audio Description Rules]}\\
Describe audio features from the perspective of the ACTIVE SPEAKER (the person who is speaking or producing vocal sound in the video).
For each category include: (1) state; (2) degree (slight / moderate / strong); (3) dynamics (no change / gradual change / fluctuating).

Required categories:
\begin{itemize}\setlength{\itemsep}{0pt}\setlength{\parskip}{0pt}
\item Speech presence: continuous speech / intermittent speech / no speech
\item Spoken content (verbatim transcription if audible): exact spoken words as heard, without interpretation, summarization, or paraphrasing
\item Voice intensity: soft / moderate / loud / variable
\item Speech rate: slow / normal / fast / changing
\item Pitch variation: flat / slightly varied / highly varied
\item Prosody dynamics: stable / gradual modulation / fluctuating
\item Articulation clarity: clear / moderately clear / unclear
\end{itemize}

\textbf{[Writing Requirements]}
\begin{itemize}\setlength{\itemsep}{0pt}\setlength{\parskip}{0pt}
\item Write in 1 unified paragraph for each modality (Scene / Facial / Audio)
\item Preserve temporal order of changes
\item Embed all attributes naturally into continuous flow
\item Keep descriptions strictly visual or acoustic and measurable
\end{itemize}
}
\end{promptboxdata}

\vspace{4pt}

\begin{promptboxdata}{Stage 2: Cognitive Appraisal (CoT)}
You are an expert in multimodal emotion analysis based on cognitive appraisal theory.
Your task is to infer the emotional state from scene, facial behavior, audio, and spoken content.

\textbf{Requirements:}
\begin{enumerate}\setlength{\itemsep}{0pt}
\item You MUST integrate evidence and appraisal inference together.
Do NOT separate ``evidence'' and ``appraisal'' into different sections.
Each appraisal dimension must be inferred together with its supporting evidence in the SAME sentence.
\item You MUST produce ONE single coherent paragraph.
No bullet points, no section headers, and no line breaks.
\item For each appraisal dimension, you MUST follow this format:
first briefly describe the supporting evidence, then explicitly state the appraisal result in parentheses.
\item You MUST include ALL six appraisal dimensions in the following order:
Goal Congruence, Coping, Accountability, Future, Arousal, Action Tendency.
\item The paragraph MUST end with a brief summary of the overall appraisal pattern and a final emotion label selected from:
happiness, sadness, anger, fear, anxiety, disgust, contempt, neutral.
\end{enumerate}
\end{promptboxdata}

\subsection{Human verification and quality filtering}
\label{app:human_verification}
\label{app:description_guideline}

To ensure that CogEmo-40K provides reliable perceptual evidence and human-aligned cognitive appraisal labels, we conduct staged human quality control after automatic generation.
Stage~1 audits description-level perception outputs; Stage~2 verifies cognition-level appraisal judgments and emotion labels. All annotators are trained with the standardized rubrics below and inspect each sample against the original video and audio.

\paragraph{Stage~1: Description-level quality assessment.}
This stage evaluates the objective multimodal descriptions generated by \texttt{Qwen3-Omni-30B-A3B}~\citep{xu2025qwen3omni}, covering Scene, Facial, and Audio.
Each sample is rated on a 1--5 scale along three criteria: \textbf{Accuracy} (whether the description matches the video and contains no hallucination), \textbf{Completeness} (whether Scene, Facial, and Audio are covered), and \textbf{Objectivity} (whether the description stays within observable evidence without emotion or psychological inference).
We randomly sample 1{,}000 descriptions and ask five researchers with backgrounds in affective computing to score each sample holistically under the rubric in \styleref{Table~\ref{tab:desc_quality_rubric}}, with priority given to Accuracy, then Completeness.
Fluent wording alone does not justify a high score; annotators must strictly separate observable description from emotion or psychological inference.

\begin{table}[htbp]
\centering
\caption{Stage~1 description-level human quality rubric (1--5).}
\label{tab:desc_quality_rubric}

\small
\renewcommand{\arraystretch}{1.12}
\setlength{\tabcolsep}{4pt}

\renewcommand{\tabularxcolumn}[1]{m{#1}}

\begin{tabularx}{\textwidth}{
>{\centering\arraybackslash}m{0.06\textwidth}
>{\centering\arraybackslash}m{0.11\textwidth}
>{\raggedright\arraybackslash}X
}
\toprule
\textbf{Score} & \textbf{Level} & \textbf{Criteria} \\
\midrule

5 & Excellent &
Fully accurate; covers Scene, Facial, and Audio; includes fine-grained cues
(e.g., gaze, mouth dynamics, prosody); fully objective with no subjective
inference; clear and unambiguous. Directly usable as high-quality data. \\

4 & Good &
Mostly accurate with only minor deviations; three modalities largely covered;
slightly coarser or fewer details; remains largely objective. Usable with
limited information loss. \\

3 & Fair &
Noticeable missing information (e.g., one modality absent); locally inaccurate
or vague; occasional mild subjective phrasing. Marginally usable but unstable
in quality. \\

2 & Poor &
Multiple inaccuracies; clear modality omissions; subjective inference or
non-compliant description. Not recommended for use. \\

1 & Bad &
Severely incorrect or inconsistent with the video; obvious hallucination;
lacks usable information. Should be discarded. \\

\bottomrule
\end{tabularx}

\end{table}

In addition to the overall rubric, annotators apply the mandatory score caps in \styleref{Table~\ref{tab:desc_deduction}}.
Records are stored as \texttt{(id, score, notes)}, where \texttt{notes} are required whenever the score is $\leq 3$.
When a sample cannot be judged reliably, annotators document the reason in \texttt{notes}.

\begin{table}[htbp]
\centering
\caption{Mandatory score caps for Stage~1 description assessment.}
\label{tab:desc_deduction}

\begin{tabular}{
>{\centering\arraybackslash}m{1.2cm}
>{\raggedright\arraybackslash}m{10cm}
}
\toprule
\textbf{Cap} & \textbf{Triggering conditions} \\
\midrule

$\leq 3$ &
Emotion words appear (e.g., angry / happy / nervous); subjective judgments
occur (e.g., ``looks angry''); any one of Scene / Facial / Audio is missing. \\

$\leq 2$ &
Multiple modalities are missing; clear inconsistency with the video;
description is severely vague or non-specific. \\

\bottomrule
\end{tabular}

\end{table}

Under this protocol, the average scores of Accuracy, Completeness, and Objectivity all exceed 4.0, indicating that Stage~1 descriptions provide reliable observable evidence for subsequent appraisal reasoning.

\paragraph{Stage~2: Cognition-level human verification.}
This stage verifies cognition-level reasoning across the six appraisal dimensions and the final emotion label. Given the substantial annotation cost, we randomly select approximately one-third of the 115{,}595 automatically generated cognition instances for human verification, without preferentially selecting samples based on their difficulty.

Before annotation, well-trained annotators receive standardized instructions and review the definitions of the six appraisal dimensions in \styleref{Table~\ref{tab:cog_check_rubric}}. For each sample, annotators first
inspect the original video and audio, considering scene context, facial behavior, body action, vocal cues, and spoken content. They then verify whether the generated reasoning for each appraisal dimension is supported by the multimodal evidence and consistent with the corresponding dimension definition. The final emotion label is additionally checked for consistency with both the
multimodal evidence and the verified appraisal reasoning.

An instance is directly retained when its emotion label agrees with human judgment and at least four of the six appraisal dimensions are judged to be consistent. We adopt four dimensions as a practical majority-based threshold: it requires broad agreement across the appraisal structure while allowing limited disagreement that may arise from the inherent subjectivity and individual variation of cognitive appraisal.

Failed instances are not discarded: human annotations are used to correct emotion labels and appraisal reasoning, regenerate evidence-grounded reasoning explanations, and undergo additional review by five affective computing researchers. This correction-based filtering yields the final CogEmo-40K of
approximately 40K high-quality instruction-tuning instances.

\begin{table}[htbp]
\centering
\caption{Definitions of the six cognitive appraisal dimensions used in Stage~2 human verification.}
\label{tab:cog_check_rubric}
\resizebox{\textwidth}{!}{%
\begin{tabular}{lp{11.5cm}}
\toprule
\textbf{Dimension} & \textbf{Definition} \\
\midrule

Goal Congruence &
Evaluates how the perceived situation relates to the subject's goals, needs,
or expectations, including whether it supports, obstructs, or has little
relevance to them. \\

Coping Potential&
Reflects the subject's perceived ability to manage, control, adapt to, or
respond effectively to the situation. \\

Accountability &
Captures how responsibility or causal attribution is assigned, including
whether the situation is attributed to the subject, another person, or no
clearly identifiable agent. \\

Future Expectancy &
Describes how the subject interprets the temporal development and expected
consequences of the situation, including anticipation, continuation, or the
absence of a salient future-oriented expectation. \\

Arousal &
Reflects the perceived level of physiological or behavioral activation
associated with the situation, as supported by observable multimodal cues such
as vocal intensity, facial dynamics, or body movement. \\

Action Tendency &
Captures the behavioral orientation implied by the appraisal, such as engaging
with, avoiding, confronting, or otherwise responding to the situation. \\

\bottomrule
\end{tabular}%
}
\end{table}

\subsection{Dataset Statistics and Comparison}
\label{app:dataset_statistics}

\begin{figure}[htbp]
    \centering
    \includegraphics[width=\linewidth]{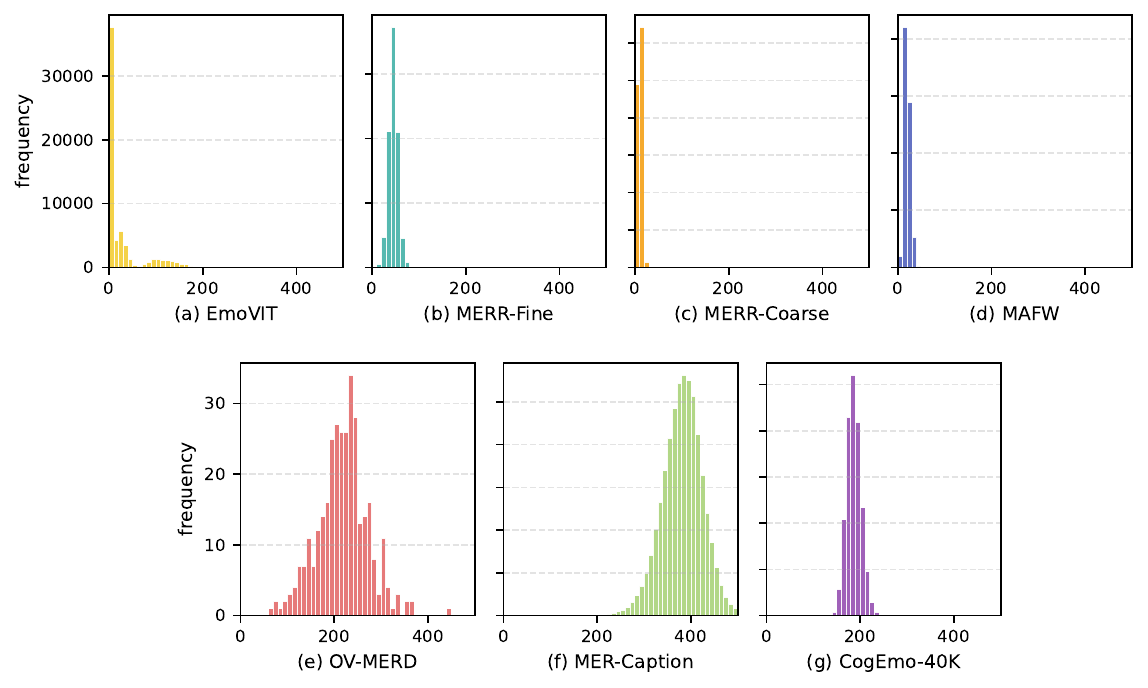}
    \caption{Distribution of description lengths (word count) across descriptive emotion datasets.}
    \label{fig:dataset_comparison}
\end{figure}

\subsubsection{Text Length Comparison}
We compare CogEmo-40K with several representative multimodal emotion datasets, including EmoVIT~\citep{xie2024emovit}, MERR-Fine~\citep{cheng2024emotionllama}, MERR-Coarse~\citep{cheng2024emotionllama}, MAFW~\citep{liu2022mafw}, OV-MERD~\citep{lian2025ov}, and MERCaption~\citep{lian2025affectgpt}, in terms of text length measured by word count. As shown in \styleref{Figure~\ref{fig:dataset_comparison}}, CogEmo-40K contains substantially longer reasoning annotations than MERR and MAFW, while remaining more concise than MERCaption. This reflects the different supervision focus of CogEmo-40K, which emphasizes structured cognitive appraisal reasoning rather than purely descriptive emotion-related content.

\subsubsection{Emotion Distribution}
\label{app:emotion_distribution}
\begin{figure*}[htbp]
    \centering
    \includegraphics[width=\textwidth]{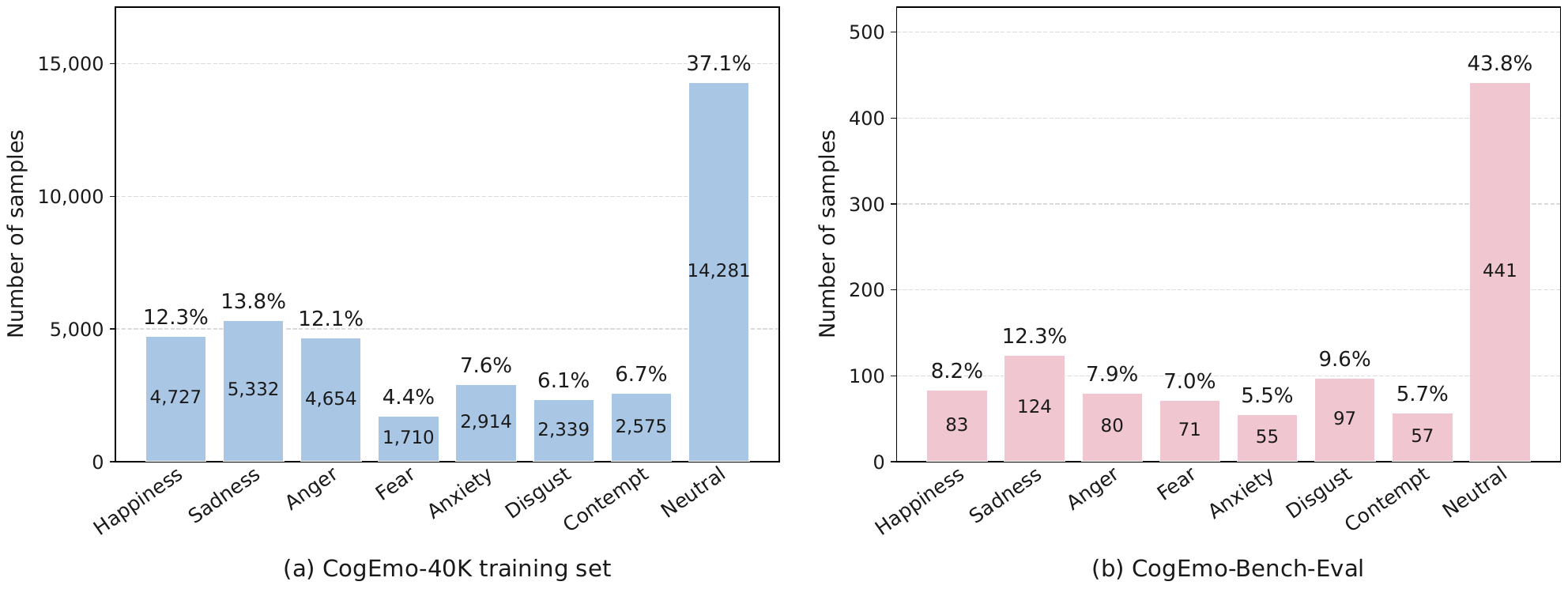}
    \caption{
    Emotion category distributions of CogEmo-40K and CogEmo-Bench-Eval.
    (a) CogEmo-40K training set ($N{=}38{,}532$).
    (b) CogEmo-Bench-Eval ($N{=}1{,}008$).
    Bars indicate the number of samples in each emotion category, with percentages annotated above.}
    \label{fig:category_dist}
\end{figure*}
As shown in \styleref{Figure~\ref{fig:category_dist}}, both CogEmo-40K and CogEmo-Bench-Eval exhibit a distribution consistent with naturalistic emotion data, where \textit{neutral} constitutes the dominant category. Beyond this overall trend, CogEmo-Bench-Eval assigns relatively larger proportions to \textit{fear} and \textit{disgust}, which are less frequent and often more challenging to distinguish. This design preserves the naturally imbalanced emotion distribution while providing stronger coverage of difficult low-frequency categories, making the benchmark more demanding for fine-grained cognition-grounded emotion reasoning.

\subsubsection{Cognitive Language Profiles}
\label{app:cognitive_profiles}

To further characterize the cognition-level supervision in CogEmo-40K, we analyze recurring language patterns across the six cognitive appraisal dimensions for each emotion category. Since the reasoning annotations are expressed in free-form natural language, we first apply the same fixed \textit{Appraisal Extractor} used in CogEmo-Bench to normalize each annotation into the six predefined appraisal dimensions. The extractor only reorganizes explicitly expressed content and does not judge correctness or introduce additional evidence.

We then group the 38,532 reasoning annotations by emotion category and appraisal dimension, and report the three most frequent semantic phrases and their occurrence rates. Percentages indicate the proportion of samples within the corresponding emotion category containing each phrase. \styleref{Table~\ref{tab:cognitive_profiles_full}} provides the full phrase-level statistics for all eight emotion categories across the six appraisal dimensions.

These statistics reveal distinct appraisal tendencies across emotions. For example, anger frequently co-occurs with goal conflict and externally directed attribution, whereas fear is more strongly associated with perceived threat, limited coping capacity, and withdrawal-related tendencies. The compact cognitive profiles presented in the main paper are semantic summaries of these recurring phrase-level patterns rather than direct copies of the raw phrases reported here.
\clearpage


\begin{table}[H]
\centering
\scriptsize
\caption{Full Cognitive Language Profiles for Eight Emotion Categories.}
\label{tab:cognitive_profiles_full}

\setlength{\tabcolsep}{2.5pt}
\renewcommand{\arraystretch}{1.02}


\textbf{Part I: Goal Congruence, Coping, and Accountability}

\vspace{2pt}

\begin{tabularx}{\textwidth}{
>{\raggedright\arraybackslash}p{0.09\textwidth}
>{\raggedright\arraybackslash}p{0.27\textwidth}
>{\raggedright\arraybackslash}X
>{\raggedright\arraybackslash}X
}
\toprule
\textbf{Emotion}
& \textbf{Goal Congruence}
& \textbf{Coping}
& \textbf{Accountability}
\\
\midrule

\textbf{Anger}
&
situation conflicts (28\%) \newline
situation involves (6\%) \newline
unfavorable evaluation (3\%)
&
able to (51\%) \newline
clear articulation (43\%) \newline
feels able (42\%)
&
another person (31\%) \newline
the other (24\%) \newline
source of (22\%)
\\

\textbf{Anxiety}
&
situation conflicts (13\%) \newline
situation involves (13\%) \newline
unfavorable evaluation (2\%)
&
clear articulation (33\%) \newline
to manage (24\%) \newline
able to (20\%)
&
source of (45\%) \newline
the source (43\%) \newline
attributed to (25\%)
\\

\textbf{Contempt}
&
situation conflicts (31\%) \newline
unfavorable evaluation (13\%) \newline
situation involves (10\%)
&
clear articulation (53\%) \newline
able to (51\%) \newline
feels able (38\%)
&
source of (33\%) \newline
the source (33\%) \newline
another person (31\%)
\\

\textbf{Disgust}
&
situation conflicts (31\%) \newline
situation involves (8\%) \newline
unfavorable evaluation (6\%)
&
clear articulation (34\%) \newline
to manage (28\%) \newline
able to (25\%)
&
the source (53\%) \newline
source of (53\%) \newline
another person (30\%)
\\

\textbf{Fear}
&
situation conflicts (9\%) \newline
situation involves (2\%)
&
perceived ability (24\%) \newline
to manage (21\%) \newline
low perceived (17\%)
&
source of (61\%) \newline
the source (58\%) \newline
attributed to (30\%)
\\

\textbf{Happiness}
&
aligns with (47\%) \newline
situation supports (35\%) \newline
situation aligns (31\%)
&
able to (50\%) \newline
feels able (41\%) \newline
clear articulation (37\%)
&
the cause (27\%) \newline
another person (15\%) \newline
the other (15\%)
\\

\textbf{Neutral}
&
aligns with (19\%) \newline
situation aligns (14\%) \newline
personal goals (7\%)
&
clear articulation (58\%) \newline
able to (55\%) \newline
to manage (41\%)
&
source of (20\%) \newline
the source (19\%) \newline
another person (16\%)
\\

\textbf{Sadness}
&
situation conflicts (26\%) \newline
situation involves (10\%) \newline
unfavorable evaluation (4\%)
&
perceived ability (20\%) \newline
relaxed jaw (18\%) \newline
clear articulation (16\%)
&
the cause (31\%) \newline
the source (22\%) \newline
source of (21\%)
\\

\bottomrule
\end{tabularx}

\vspace{4pt}


\textbf{Part II: Future, Arousal, and Action Tendency}

\vspace{2pt}

\begin{tabularx}{\textwidth}{
>{\raggedright\arraybackslash}p{0.09\textwidth}
>{\raggedright\arraybackslash}p{0.27\textwidth}
>{\raggedright\arraybackslash}X
>{\raggedright\arraybackslash}X
}
\toprule
\textbf{Emotion}
& \textbf{Future}
& \textbf{Arousal}
& \textbf{Action Tendency}
\\
\midrule

\textbf{Anger}
&
the present (21\%) \newline
a present (17\%) \newline
distant possibility (13\%)
&
physiological activation (32\%) \newline
facial tension (31\%) \newline
point to (28\%)
&
inclination to (59\%) \newline
behavioral inclination (48\%) \newline
a behavioral (47\%)
\\

\textbf{Anxiety}
&
concern about (19\%) \newline
points to (18\%) \newline
the present (15\%)
&
activation rather (34\%) \newline
facial mobility (33\%) \newline
low facial (32\%)
&
inclination to (53\%) \newline
behavioral inclination (34\%) \newline
a behavioral (33\%)
\\

\textbf{Contempt}
&
the present (23\%) \newline
a present (21\%) \newline
future threat (13\%)
&
facial mobility (58\%) \newline
low facial (55\%) \newline
activation rather (54\%)
&
inclination to (48\%) \newline
behavioral inclination (27\%) \newline
a behavioral (27\%)
\\

\textbf{Disgust}
&
the present (25\%) \newline
tied to (18\%) \newline
a present (17\%)
&
point to (39\%) \newline
facial mobility (36\%) \newline
low facial (31\%)
&
inclination to (56\%) \newline
behavioral inclination (48\%) \newline
a behavioral (47\%)
\\

\textbf{Fear}
&
concern about (25\%) \newline
what may (21\%) \newline
about what (21\%)
&
physiological activation (31\%) \newline
point to (17\%) \newline
facial tension (16\%)
&
inclination to (40\%) \newline
behavioral inclination (36\%) \newline
a behavioral (33\%)
\\

\textbf{Happiness}
&
the present (28\%) \newline
points to (17\%) \newline
a present (12\%)
&
facial mobility (57\%) \newline
point to (41\%) \newline
low facial (41\%)
&
inclination to (53\%) \newline
inclination toward (43\%) \newline
behavioral inclination (39\%)
\\

\textbf{Neutral}
&
points to (21\%) \newline
the present (20\%) \newline
a present (13\%)
&
low facial (69\%) \newline
facial mobility (69\%) \newline
point to (46\%)
&
inclination to (30\%) \newline
orientation toward (25\%) \newline
toward the (24\%)
\\

\textbf{Sadness}
&
the present (21\%) \newline
future threat (16\%) \newline
a present (14\%)
&
activation rather (37\%) \newline
point to (35\%) \newline
facial mobility (29\%)
&
inclination to (61\%) \newline
behavioral inclination (47\%) \newline
a behavioral (46\%)
\\

\bottomrule
\end{tabularx}

\end{table}

\clearpage
\section{IMPLEMENTATION DETAILS}
\label{app:implementation_details}

\subsection{Details of CogEmo-MoE Model}
CogEmo-MoE extends the MoE-LLaVA~\citep{lin2026moe} framework to support joint video, audio, and text modeling, while maintaining a compact model scale of approximately 3.1B parameters. It adopts a Qwen-1.8B~\citep{bai2023qwen} language model with CLIP ViT-L/14~\citep{radford2021learning} and CLAP~\citep{elizalde2023clap} as the visual and audio encoders, respectively, and uses two-layer GELU MLP projectors for modality alignment. The model is trained progressively for multimodal alignment, perception enhancement, and cognitive appraisal specialization using DeepSpeed ZeRO-2 with bfloat16, and a maximum sequence length of 2048 tokens. All experiments are conducted on two NVIDIA A100 80GB GPUs. Detailed configurations are summarized in \styleref{Table~\ref{tab:cogemo_moe_training}}.

\begin{table}[htbp]
\centering
\caption{Training configuration of CogEmo‑MoE across three stages.}
\label{tab:cogemo_moe_training}
\resizebox{\textwidth}{!}{%
\begin{tabular}{lccc}
\toprule
\textbf{Configuration} & \textbf{Stage 1} & \textbf{Stage 2} & \textbf{Stage 3} \\
\midrule
Trainable modules & Projectors & Full model & MoE FFN + router \\
\midrule
MoE enabled & -- & -- & Sparse (4 experts, top‑2) \\
Capacity factor & -- & -- & 1.5 \\
Router aux.\ loss $\lambda$ & -- & -- & 0.01 \\
\midrule
Epochs & 4 & 1 & 6 \\
Learning rate & $1\times10^{-3}$ & $2\times10^{-5}$ & $1\times10^{-5}$ \\
Warmup ratio & 0.03 & 0.03 & 0.05 \\
LR scheduler & cosine & cosine & cosine \\
Batch size (per device) & 8 & 8 & 4 \\
Gradient accumulation & 1 & 2 & 1 \\
\midrule
Max sequence length & 2048 & 2048 & 2048 \\
Precision & bf16 & bf16 & bf16 \\
DeepSpeed strategy & DeepSpeed ZeRO‑2 & DeepSpeed ZeRO‑2 & DeepSpeed ZeRO‑2 \\
\midrule
\end{tabular}%
}
\end{table}


\subsubsection{Training prompts for each stage}
\label{app:instruction}

Each training stage uses a task-specific user instruction and multimodal inputs consisting of video frames and audio. The placeholders \texttt{<frames>} and \texttt{<audio>} indicate the positions where visual and acoustic features encoded by CLIP~\citep{radford2021learning} and CLAP~\citep{elizalde2023clap} are inserted, respectively. The supervision targets differ across stages: Stage~1 uses coarse multimodal descriptions from MERR-Coarse~\citep{cheng2024emotionllama}, Stage~2 uses perception-level descriptions from CogEmo-Desc, and Stage~3 uses cognition-level appraisal reasoning from CogEmo-40K.

\paragraph{Stage~1: Multimodal Alignment.}
The user instruction elicits a concise, free-form description of how the subject looks and sounds, providing a lightweight alignment signal for the multimodal projectors.

\begin{promptboxtrain}{Stage 1: Multimodal Alignment (user instruction)}
Here is the video: \texttt{<frames>}\\
Here is the audio: \texttt{<audio>}\\[0.4em]
Describe how the person looks and sounds in the video.
\end{promptboxtrain}

\paragraph{Stage~2: Perception enhancement.}
The instruction shifts toward structured, high-density perceptual description of the scene, facial behavior, and acoustic characteristics.
The target response follows the emotion-free perception format produced by the dataset Stage~1 generation prompt.

\begin{promptboxtrain}{Stage 2: Perception Enhancement (user instruction)}
Here is the video: \texttt{<frames>}\\
Here is the audio: \texttt{<audio>}\\[0.4em]
Please briefly describe the video scene, and provide a detailed description of the main subject's facial details and the speaker's acoustic characteristics.
\end{promptboxtrain}

\paragraph{Stage~3: Cognitive appraisal tuning.}
The instruction asks the model to perform cognitive appraisal reasoning over all available modalities and output a final discrete emotion label.
The target response follows the appraisal chain-of-thought format produced by the dataset Stage~2 generation prompt.

\begin{promptboxtrain}{Stage 3: Cognitive Appraisal Tuning (user instruction)}
Here is the video: \texttt{<frames>}\\
Here is the audio: \texttt{<audio>}\\[0.4em]
Please infer the most likely emotion category from multimodal cues by reasoning through six cognitive appraisal dimensions: Congruence, Coping, Accountability, Future, Arousal, and Action Tendency. The final emotion should be selected from happiness, sadness, anger, fear, anxiety, disgust, contempt, and neutral.
\end{promptboxtrain}

\subsection{Model Scale and Performance Comparison}
\label{app:model_scale}

\styleref{Table~\ref{tab:performance_size}} compares model scale and performance on CogEmo-Bench. While most multimodal emotion LLMs use backbones of at least 7B parameters, the 3.1B CogEmo-MoE achieves the best AEQS, accuracy, and Macro-F1, outperforming the strongest baselines by 14.30, 8.33, and 4.98 points, respectively. This demonstrates that its performance gains arise from effective cognitive modeling rather than increased model scale.

\begin{table*}[htbp]
\centering
\small
\caption{Performance--Scale Comparison on CogEmo-Bench.}
\label{tab:performance_size}

\vspace{-2pt}
\setlength{\tabcolsep}{9pt}
\renewcommand{\arraystretch}{1.08}

\begin{tabular}{lcccc}
\toprule
\textbf{Model}
& \textbf{Parameters}
& \textbf{AEQS}
& \textbf{Acc.}
& \textbf{Macro-F1}
\\
\midrule


\rowcolor{groupteal}
\multicolumn{5}{c}{\textit{Open-source MLLMs}}
\\

OneLLM~\citep{han2024onellm}
& 7B
& 16.15
& 30.16
& 26.03
\\

SECap~\citep{xu2024secap}
& 7B
& 12.26
& 30.85
& 21.25
\\

Qwen2-Audio~\citep{chu2024qwen2audio}
& 7B
& 25.12
& 43.55
& 32.68
\\

SALMONN~\citep{tang2024salmonn}
& 7B
& 23.87
& 43.69
& 33.43
\\

Otter~\citep{li2025otter}
& 7B
& 10.32
& 36.61
& 24.69
\\

LLaMA-VID~\citep{li2024llama}
& 7B
& 26.60
& 36.71
& 31.84
\\

VideoChat-Flash~\citep{li2025videochat}
& 7B
& 2.69
& 55.36
& 39.11
\\

Chat-UniVi~\citep{jin2024chat}
& 7B
& 32.28
& 34.03
& 29.86
\\

mPLUG-Owl3~\citep{ye2025mplug}
& 7B
& 32.07
& 45.54
& 34.72
\\

Qwen3.8-Max~\citep{team2026qwen3}
& 2.4T
& 49.50
& 57.04
& 46.29
\\

Kimi-K3~\citep{kimi2026k3}
& 2.8T (104B active)
& 48.70
& 46.13
& 39.04
\\

PandaGPT~\citep{su2023pandagpt}
& 7B
& 25.54
& 25.79
& 18.78
\\

Emotion-LLaMA~\citep{cheng2024emotionllama}
& 7B
& 20.27
& 43.75
& 25.90
\\

AffectGPT~\citep{lian2025affectgpt}
& 7B
& 32.83
& 42.76
& 31.90
\\

Qwen3-Omni~\citep{xu2025qwen3omni}
& 30B (3B active)
& 46.12
& 63.10
& 52.19
\\

\midrule


\rowcolor{groupteal}
\multicolumn{5}{c}{\textit{Closed-source MLLMs}}
\\

GPT-5.2~\citep{openai2025gpt52}
& ——
& 56.30
& 46.33
& 36.92
\\

GPT-5.6-Luna~\citep{openai2026gpt56}
& ——
& 48.27
& 51.59
& 43.46
\\

Grok-4.3~\citep{xai2026grok43}
& ——
& 45.55
& 46.83
& 39.79
\\

Gemini-3.6-Flash~\citep{google2026gemini36flash}
& -.
& 36.30
& 59.13
& 45.99
\\

\midrule


\rowcolor{groupteal}
\multicolumn{5}{c}{\textit{Our Model}}
\\

\textbf{CogEmo-MoE}
& \textbf{3.1B}
& \textbf{70.60}
& \textbf{71.43}
& \textbf{57.17}
\\

\bottomrule
\end{tabular}
\end{table*}

\clearpage
\section{Details of CogEmo-Bench}
\label{app:cogemo_bench}

CogEmo-Bench is a held-out evaluation set providing over 6,000 human-verified dimension-level appraisal references across six cognitive appraisal dimensions, under the same eight-class emotion space as CogEmo-40K. It is used to assess both cognitive appraisal evidence quality and final emotion recognition under a unified protocol. Its label distribution is shown beside the training set in \styleref{Figure~\ref{fig:category_dist}}. Consistent with that training panel, neutral remains dominant (43.8\%), while the remaining seven categories range from 5.5\% (anxiety) to 12.3\% (sadness). Relative to CogEmo-40K, disgust and fear occupy a larger share of the evaluation split, which helps stress-test appraisal reasoning under less frequent negative emotions. The remainder of this section details the evaluated models, prompting templates, appraisal extractor, and metric computation.

\subsection{Model setup}
\label{app:bench_model_setup}

We evaluate all models reported in \styleref{Table~\ref{tab:cogemo_bench}} on the CogEmo-Bench test set.
Each model is loaded from its official release and evaluated with the decoding configuration recommended by the original implementation.
\styleref{Table~\ref{tab:bench_model_setup}} summarizes the model category, supported modalities, and the corresponding evaluation-prompt template.
For open-source models, we follow the default checkpoints and preprocessing pipelines released by the original authors.
For closed-source models, we access the latest available API endpoints at the time of evaluation.
CogEmo-MoE is evaluated under three modality settings (A+T, V+T, and A+V+T) to match the ablations in the main benchmark table.
General-purpose MLLMs (Gen.) are prompted for cognitive appraisal reasoning with modality-specific templates; emotion-specialized MLLMs (Spec.) follow their original prompting protocols.

\begin{table*}[htbp]
\centering
\caption{
\textbf{Evaluation setup for models on CogEmo-Bench.} Gen. and Spec. denote general-purpose MLLMs and emotion-specialized models, respectively. A, V, and T denote audio, visual, and textual modalities. Evaluation prompt templates are provided in \styleref{Appendix~\ref{app:bench_model_prompt}}.}
\label{tab:bench_model_setup}
\setlength{\tabcolsep}{5pt}
\renewcommand{\arraystretch}{1.08}
\resizebox{\textwidth}{!}{%
\begin{tabular}{llccc|l}
\toprule
\textbf{Model} & \textbf{Type} & \textbf{A} & \textbf{V} & \textbf{T} & \textbf{Prompt Template} \\
\midrule
\rowcolor{groupteal}
\multicolumn{6}{c}{\textit{Open-source MLLMs}} \\
OneLLM~\citep{han2024onellm} & Gen. & $\checkmark$ & $\times$ & $\checkmark$ & AT (\styleref{Appendix~\ref{app:bench_prompt_at}}) \\
SECap~\citep{xu2024secap} & Spec. & $\checkmark$ & $\times$ & $\checkmark$ & SECap (\styleref{Appendix~\ref{app:bench_prompt_secap}}) \\
Qwen2-Audio~\citep{chu2024qwen2audio} & Gen. & $\checkmark$ & $\times$ & $\checkmark$ & AT (\styleref{Appendix~\ref{app:bench_prompt_at}}) \\
SALMONN~\citep{tang2024salmonn} & Gen. & $\checkmark$ & $\times$ & $\checkmark$ & AT (\styleref{Appendix~\ref{app:bench_prompt_at}}) \\
Otter~\citep{li2025otter} & Gen. & $\times$ & $\checkmark$ & $\checkmark$ & VT (\styleref{Appendix~\ref{app:bench_prompt_vt}}) \\
LLaMA-VID~\citep{li2024llama} & Gen. & $\times$ & $\checkmark$ & $\checkmark$ & VT (\styleref{Appendix~\ref{app:bench_prompt_vt}}) \\
VideoChat-Flash~\citep{li2025videochat} & Gen. & $\times$ & $\checkmark$ & $\checkmark$ & VT (\styleref{Appendix~\ref{app:bench_prompt_vt}}) \\
Chat-UniVi~\citep{jin2024chat} & Gen. & $\times$ & $\checkmark$ & $\checkmark$ & VT (\styleref{Appendix~\ref{app:bench_prompt_vt}}) \\
mPLUG-Owl3~\citep{ye2025mplug} & Gen. & $\times$ & $\checkmark$ & $\checkmark$ & VT (\styleref{Appendix~\ref{app:bench_prompt_vt}}) \\
Qwen3.8-Max~\citep{team2026qwen3} & Gen. & $\times$ & $\checkmark$ & $\checkmark$ & VT (\styleref{Appendix~\ref{app:bench_prompt_vt}}) \\
Kimi-K3~\citep{kimi2026k3} & Gen. & $\times$ & $\checkmark$ & $\checkmark$ & VT (\styleref{Appendix~\ref{app:bench_prompt_vt}}) \\
PandaGPT~\citep{su2023pandagpt} & Gen. & $\checkmark$ & $\checkmark$ & $\checkmark$ & AVT (\styleref{Appendix~\ref{app:bench_prompt_avt}}) \\
Emotion-LLaMA~\citep{cheng2024emotionllama} & Spec. & $\checkmark$ & $\checkmark$ & $\checkmark$ & Emotion-LLaMA (\styleref{Appendix~\ref{app:bench_prompt_emotionllama}}) \\
AffectGPT~\citep{lian2025affectgpt} & Spec. & $\checkmark$ & $\checkmark$ & $\checkmark$ & AffectGPT (\styleref{Appendix~\ref{app:bench_prompt_affectgpt}}) \\
Qwen3-Omni~\citep{xu2025qwen3omni} & Gen. & $\checkmark$ & $\checkmark$ & $\checkmark$ & AVT (\styleref{Appendix~\ref{app:bench_prompt_avt}}) \\
\midrule
\rowcolor{groupteal}
\multicolumn{6}{c}{\textit{Closed-source MLLMs}} \\
GPT-5.2~\citep{openai2025gpt52} & Gen. & $\times$ & $\checkmark$ & $\checkmark$ & VT (\styleref{Appendix~\ref{app:bench_prompt_vt}}) \\
GPT-5.6-Luna~\citep{openai2026gpt56} & Gen. & $\times$ & $\checkmark$ & $\checkmark$ & VT (\styleref{Appendix~\ref{app:bench_prompt_vt}}) \\
Grok-4.3~\citep{xai2026grok43} & Gen. & $\times$ & $\checkmark$ & $\checkmark$ & VT (\styleref{Appendix~\ref{app:bench_prompt_vt}}) \\
Gemini-3.6-Flash~\citep{google2026gemini36flash} & Gen. & $\checkmark$ & $\checkmark$ & $\checkmark$ & AVT (\styleref{Appendix~\ref{app:bench_prompt_avt}}) \\
\midrule
\rowcolor{groupteal}
\multicolumn{6}{c}{\textit{Our CogEmo Models}} \\
CogEmo-MoE (A+T) & Spec. & $\checkmark$ & $\times$ & $\checkmark$ & AT (\styleref{Appendix~\ref{app:bench_prompt_at}}) \\
CogEmo-MoE (V+T) & Spec. & $\times$ & $\checkmark$ & $\checkmark$ & VT (\styleref{Appendix~\ref{app:bench_prompt_vt}}) \\
CogEmo-MoE (A+V+T) & Spec. & $\checkmark$ & $\checkmark$ & $\checkmark$ & AVT (\styleref{Appendix~\ref{app:bench_prompt_avt}}) \\
\bottomrule
\end{tabular}%
}
\end{table*}

\subsection{Evaluation Prompt of Different Models}
\label{app:bench_model_prompt}

All models are evaluated on the same CogEmo-Bench instances using the input modalities supported by each model. Specifically, video frames, audio waveforms, and subtitle transcripts are provided whenever the corresponding modality is supported, while unsupported modalities are omitted. General-purpose MLLMs are prompted to perform cognitive appraisal reasoning and produce a final emotion prediction, whereas emotion-specialized models follow their original prompting protocols to preserve their intended inference behavior.

Accordingly, we use two types of evaluation prompts: model-specific prompts for emotion-specialized models and modality-specific appraisal prompts for general-purpose MLLMs, as detailed below.

\paragraph{Emotion-specialized models.}
\label{app:bench_prompt_specialist}

\begin{promptboxbench}{AffectGPT (user instruction)}
\label{app:bench_prompt_affectgpt}
Please infer the person's emotional state and provide your reasoning process.
\end{promptboxbench}

\begin{promptboxbench}{Emotion-LLaMA (user instruction)}
\label{app:bench_prompt_emotionllama}
[reason] Please analyze all clues and reason out the emotional label.
\end{promptboxbench}

\begin{promptboxbench}{SECap (user instruction)}
\label{app:bench_prompt_secap}
Please describe the speaker's emotional state in one sentence.
\end{promptboxbench}
SECap follows its original protocol of using a concise emotion-description prompt.
The original implementation provides multiple Chinese prompt paraphrases with equivalent meaning; we use the single English sentence above for consistency across models.

\paragraph{General-purpose MLLMs.}
\label{app:bench_prompt_general}

Depending on the supported modalities (\styleref{Table~\ref{tab:bench_model_setup}}), we use one of the following three prompt templates.

\begin{promptboxbench}{A+V+T prompt (AVT)}
\label{app:bench_prompt_avt}
Given the multimodal input, analyze the person's emotional state and explain why the person may feel this way.
Ground your reasoning in the available visual, audio, textual, and contextual evidence.
Finally, identify the most likely emotion from:
happiness, sadness, anger, fear, anxiety, disgust, contempt, neutral.
\end{promptboxbench}

\begin{promptboxbench}{A+T prompt (AT)}
\label{app:bench_prompt_at}
Given the multimodal input, analyze the person's emotional state and explain why the person may feel this way.
Ground your reasoning in the available audio, textual, and contextual evidence.
Finally, identify the most likely emotion from:
happiness, sadness, anger, fear, anxiety, disgust, contempt, neutral.
\end{promptboxbench}

\begin{promptboxbench}{V+T prompt (VT)}
\label{app:bench_prompt_vt}
Given the multimodal input, analyze the person's emotional state and explain why the person may feel this way.
Ground your reasoning in the available visual, textual, and contextual evidence.
Finally, identify the most likely emotion from:
happiness, sadness, anger, fear, anxiety, disgust, contempt, neutral.
\end{promptboxbench}

\paragraph{Appraisal Extractor.}
\label{app:bench_parser_prompt}

After each model generates a free-form response, we apply a fixed extraction prompt implemented with \texttt{Qwen3.5-9B}~\citep{qwen2026qwen35} to obtain six appraisal-specific evidence units and the final emotion label.
The extractor only reorganizes information explicitly present in the original response; it does not add new evidence or modify the model's reasoning.
If a dimension is absent in the response, its extracted evidence is left empty.
All evaluated models share the same extractor and prompt to ensure a consistent evaluation protocol.

\begin{promptboxbench}{Appraisal Extractor (user instruction)}
You are an appraisal extractor.
Given a model-generated emotion reasoning response, extract the supporting evidence and appraisal judgment for each of the following dimensions if explicitly present:
Goal Congruence, Coping, Accountability, Future, Arousal, and Action Tendency.
Also extract the final emotion label.
Do not add new evidence or modify the original reasoning.
If a dimension is missing, leave it empty.
\end{promptboxbench}

\subsection{Metric Computation Details}
\label{app:bench_metrics}

CogEmo-Bench jointly evaluates cognitive appraisal reasoning and final emotion recognition. The former is measured by the \textbf{Appraisal Evidence Quality Score (AEQS)}, while the latter is evaluated using \textbf{Accuracy} and \textbf{Macro-F1}. The formal definitions are introduced in \styleref{Section~\ref{sec:benchmark_metrics}}; here we provide the corresponding implementation details.

\paragraph{Evaluation set.}
CogEmo-Bench-Eval is a human-verified multimodal evaluation set under the same eight-class emotion space as CogEmo-40K: \{\textit{happiness}, \textit{sadness}, \textit{anger}, \textit{fear}, \textit{anxiety}, \textit{disgust}, \textit{contempt}, \textit{neutral}\}. Each instance provides human-verified cognitive appraisal references across six dimensions together with a final emotion label.

\paragraph{Response normalization.}
For each instance, the evaluated model first generates a free-form response using the corresponding evaluation prompt described in \styleref{Appendix~\ref{app:bench_model_prompt}}. The response is then processed by the same fixed \textit{Appraisal Extractor} (\styleref{Appendix~\ref{app:bench_parser_prompt}}), which reorganizes the explicitly expressed content into six appraisal dimensions and a final emotion prediction. The extractor does not judge correctness or introduce additional evidence. Appraisal information not expressed in the original response is retained as empty.

\paragraph{Appraisal Evidence Quality Score.}
For each sample $i$ and appraisal dimension $d$, we compare the extracted model evidence $\hat{e}_{i,d}$ with the corresponding human-verified reference $e_{i,d}$. Both are encoded using the same fixed \texttt{Qwen3-Embedding} model, and their semantic similarity is computed by cosine similarity as defined in \styleref{Eq.~\eqref{eq:appraisal_similarity}}.

When only one side contains appraisal evidence, the similarity score is set to zero, penalizing either missing or unsupported evidence. When both sides are empty, the sample is excluded from the corresponding dimension because no evidence is available for semantic comparison. We first average the valid sample-level similarities within each appraisal dimension and then average the six dimension-level scores to obtain AEQS according to \styleref{Eq.~\eqref{eq:aeqs}}. For readability, the final AEQS is multiplied by 100 when reported.

\textbf{Emotion Recognition.}
For final emotion prediction, we report \textbf{Accuracy} and
\textbf{Macro-F1}. Accuracy measures the proportion of correctly predicted
emotion labels, while Macro-F1 equally weights the class-wise F1 scores across
all emotion categories:

\begin{equation}
\mathrm{Acc.}
=
\frac{1}{N}\sum\nolimits_{i=1}^{N}\mathbb{I}(\hat{y}_i=y_i),
\qquad
\mathrm{Macro\text{-}F1}
=
\frac{1}{C}\sum\nolimits_{c=1}^{C}
\frac{2P_cR_c}{P_c+R_c}.
\label{eq:emotion_metrics}
\end{equation}

where $\hat{y}_i$ and $y_i$ denote the predicted and ground-truth emotion
labels for sample $i$, respectively; $P_c$ and $R_c$ denote the precision and
recall for class $c$; $N$ is the number of evaluation samples; and $C=8$ is
the number of emotion categories.

\paragraph{Unified evaluation.}
All evaluated models share the same Appraisal Extractor, embedding model, emotion label space, and metric computation procedure. This unified pipeline minimizes evaluation differences introduced by heterogeneous response formats and enables consistent comparison of both appraisal reasoning and emotion recognition across models.

\clearpage
\section{ADDITIONAL EXPERIMENTAL RESULTS}
\label{app:additional_results}

\subsection{Cognitive Reasoning--Emotion Alignment}
\label{app:5fold}
To examine whether the cognitive appraisal reasoning in CogEmo-40K is aligned with the corresponding final emotion, we conduct a 5-fold cross-validation experiment on the \textbf{intersection subset of CogEmo-40K and MERCaption+} (11{,}770 samples)~\citep{lian2025affectgpt}. Specifically, we use the cognitive reasoning annotations as representations and formulate emotion prediction as an 8-class classification task. This experiment is intended as a diagnostic analysis of the relationship between intermediate cognitive reasoning and the resulting emotion label, rather than an evaluation of a standalone text classifier. As shown in \styleref{Table~\ref{tab:5fold_reason}}, CogEmo-40K reasoning representations exhibit strong emotion separability, suggesting that the generated appraisal chains encode emotion-relevant cognitive information that is closely aligned with the final emotion labels.
\begin{table}[htbp]
\centering
\caption{
5-fold cross-validation results on the text-reason branch of the \textbf{CogEmo-40K \& MERCaption+~\citep{lian2025affectgpt} intersection subset}.
Accuracy, Weighted-F1, and Macro-F1 are reported in \%.
The best average is shown in bold.
}
\label{tab:5fold_reason}
\setlength{\tabcolsep}{8pt}
\renewcommand{\arraystretch}{1.12}

\begin{tabular}{lccccc>{\bfseries}c}
\toprule
\textbf{Metric}
& \textbf{Fold 1}
& \textbf{Fold 2}
& \textbf{Fold 3}
& \textbf{Fold 4}
& \textbf{Fold 5}
& \textbf{Avg$\pm$Std}
\\
\midrule
\rowcolor{gray!12}
Accuracy & 92.44 & 92.57 & 91.59 & 92.44 & 92.40 & 92.29$\pm$0.35 \\
Weighted-F1 & 92.43 & 92.57 & 91.50 & 92.48 & 92.31 & 92.26$\pm$0.39 \\
\rowcolor{gray!12}
Macro-F1 & 91.44 & 91.99 & 90.27 & 91.72 & 91.23 & 91.33$\pm$0.59 \\
\bottomrule
\end{tabular}
\end{table}

\paragraph{Training Details.}
The 5-fold setup follows a fixed, pre-generated split (seed 42, 80\%/20\% per fold) on the intersection subset described above, ensuring full reproducibility. We list the key training configurations below:
\begin{enumerate}
    \item \textbf{Backbone:} XLM-RoBERTa-base.
    \item \textbf{Optimizer:} AdamW (HuggingFace Trainer default).
    \item \textbf{Initial learning rate:} $2\times10^{-5}$ with a linear decay schedule and no warmup.
    \item \textbf{Weight decay:} 0.01.
    \item \textbf{Batch size:} 32 per device.
    \item \textbf{Epochs:} 15.
    \item \textbf{Maximum sequence length:} 512.
    \item \textbf{Precision:} Full precision (FP32); no gradient accumulation.
    \item \textbf{Hardware:} Single NVIDIA L40 (48GB).
    \item \textbf{Data preprocessing:} A rule-based label-leakage stripping is applied to remove final explicit emotion-conclusion sentences (e.g., ``therefore the emotion is anger''), forcing the model to rely on the actual reasoning content.
\end{enumerate}

In addition to the classification results, we perform a feature-space analysis by extracting mean-pooled XLM-RoBERTa embeddings of the reasoning text, applying PCA, and computing the Silhouette score on 5{,}000 sampled instances. The resulting cluster separability further supports the quality of the cognitive reasoning annotations.

%

\subsection{Cross-Domain Generalization Experiments}
\label{app:cross_domain}

To evaluate the generalization ability of cognitive appraisal-guided training, we conduct a controlled comparison between our automatically generated cognition-grounded supervision and existing emotion description supervision. Specifically, we use the same 115{,}595 source samples as MERCaption~\citep{lian2025affectgpt} and generate cognitive appraisal annotations under our perception-to-appraisal paradigm without human refinement. In contrast, MERCaption provides automatically generated free-form affective descriptions on the same source set. This setup controls for both data source and data scale, allowing us to more directly examine the effect of the supervision paradigm.

All models are trained under the same AffectGPT~\citep{lian2025affectgpt} framework and evaluated following the MERBench protocol~\citep{lian2025affectgpt}, covering nine emotion datasets. The corresponding zero-shot results are reported in \styleref{Table~\ref{tab:cross_domain_part1}} for Part I (MER2023--CMU-MOSI) and \styleref{Table~\ref{tab:cross_domain_part2}} for Part II (CMU-MOSEI--OV-MERD+), covering both General Instruction and Emotion Description tasks. A detailed description of MERBench is provided in Appendix~\ref{app:merbench}.

\begin{table}[!ht] 
\centering
\setlength{\abovecaptionskip}{2pt}
\setlength{\belowcaptionskip}{2pt}
\caption{
Cross-domain generalization results on \textbf{General Instruction} and \textbf{Emotion Description} datasets (Part I: MER2023--CMU-MOSI). All models are trained on AffectGPT and evaluated on MERBench. Top: General Instruction models. Bottom: Emotion Description models. CogEmo-40K is highlighted in blue.
}
\label{tab:cross_domain_part1}
\small 
\setlength{\tabcolsep}{3pt}

\resizebox{\textwidth}{!}{
\begin{tabular}{lcccccc}
\toprule
\textbf{Dataset} & \textbf{Filtering} & \textbf{MER2023} & \textbf{MER2024} & \textbf{MELD} & \textbf{IEMOCAP} & \textbf{CMU-MOSI} \\
\midrule
MiniGPT4 & $\times$ & 11.56 & 12.91 & 18.89 & 16.06 & 53.57 \\
MiniGPT4 & $\checkmark$ & 17.57 & 16.65 & 22.60 & 30.18 & 52.58 \\
VideoChat & $\times$ & 24.87 & 22.42 & 21.56 & 32.91 & 50.13 \\
VideoChat & $\checkmark$ & 27.70 & 24.73 & 27.66 & 39.46 & 45.45 \\
LLaVA & $\times$ & 42.21 & 41.54 & 32.97 & 49.96 & 54.48 \\
LLaVA & $\checkmark$ & 41.56 & 42.30 & 32.61 & 46.21 & 52.82 \\
WavCaps & $\times$ & 5.75 & 7.71 & 4.35 & 4.99 & 45.59 \\
WavCaps & $\checkmark$ & 23.72 & 26.97 & 23.39 & 27.30 & 54.67 \\
\midrule
EmoVIT & -- & 39.31 & 50.24 & 32.36 & 48.24 & 53.40 \\
MAFW & -- & 52.67 & 55.99 & 40.85 & 57.60 & 66.11 \\
MERR-Coarse & -- & 35.34 & 36.60 & 29.37 & 36.94 & 65.10 \\
MERR-Fine & -- & 69.00 & 72.84 & 47.38 & 54.49 & 66.21 \\
MERCaption & -- & 72.12 & \textbf{74.21} & 54.69 & 56.41 & 75.10 \\
\rowcolor{blue!15}
\textbf{CogEmo-40K (Ours)} & \textbf{--} & \textbf{74.01} & 72.66 & \textbf{57.35} & \textbf{57.53} & \textbf{79.15} \\
\bottomrule
\end{tabular}
} 
\end{table}

\begin{table}[!ht] 
\centering
\setlength{\abovecaptionskip}{2pt}
\setlength{\belowcaptionskip}{2pt}
\caption{
Cross-domain generalization results on \textbf{General Instruction} and \textbf{Emotion Description} datasets (Part II: CMU-MOSEI--OV-MERD+). All models are trained on AffectGPT and evaluated on MERBench. Top: General Instruction models. Bottom: Emotion Description models. CogEmo-40K is highlighted in blue. Mean is computed across all nine datasets.
}
\label{tab:cross_domain_part2}
\small 
\setlength{\tabcolsep}{3pt}

\resizebox{\textwidth}{!}{
\begin{tabular}{lcccccc}
\toprule
\textbf{Dataset} & \textbf{Filtering} & \textbf{CMU-MOSEI} & \textbf{CH-SIMS} & \textbf{CH-SIMS v2} & \textbf{OV-MERD+} & \textbf{Mean} \\
\midrule
MiniGPT4 & $\times$ & 45.98 & 57.66 & 55.16 & 13.86 & 31.74 \\
MiniGPT4 & $\checkmark$ & 56.50 & 52.36 & 51.19 & 20.16 & 35.53 \\
VideoChat & $\times$ & 56.17 & 50.07 & 51.71 & 24.56 & 37.16 \\
VideoChat & $\checkmark$ & 56.86 & 43.68 & 47.05 & 26.09 & 37.63 \\
LLaVA & $\times$ & 56.42 & 52.04 & 54.80 & 35.75 & 46.69 \\
LLaVA & $\checkmark$ & 57.72 & 52.78 & 53.44 & 36.96 & 46.27 \\
WavCaps & $\times$ & 22.76 & 53.04 & 45.68 & 4.95 & 21.65 \\
WavCaps & $\checkmark$ & 49.54 & 58.12 & 55.93 & 21.57 & 37.91 \\
\midrule
EmoVIT & -- & 61.53 & 69.72 & 66.53 & 38.09 & 51.05 \\
MAFW & -- & 62.27 & 75.20 & 70.02 & 42.75 & 58.16 \\
MERR-Coarse & -- & 63.27 & 75.12 & 73.76 & 33.18 & 49.85 \\
MERR-Fine & -- & 60.03 & 79.90 & 78.54 & 52.56 & 64.55 \\
MERCaption & -- & \textbf{70.97} & 80.21 & \textbf{78.06} & \textbf{58.42} & 68.91 \\
\rowcolor{blue!15}
\textbf{CogEmo-40K (Ours)} & \textbf{--} & 69.75 & \textbf{83.01} & 77.94 & 58.16 & \textbf{69.95} \\
\bottomrule
\end{tabular}
} 
\end{table}

\paragraph{Discussion.}
Compared with the strongest emotion description baseline, \textbf{MERCaption}~\citep{lian2025affectgpt}, \textbf{CogEmo-40K} improves performance on most benchmarks, including MER2023~\citep{lian2023mer}, MER2024~\citep{lian2024mer}, MELD~\citep{poria-etal-2019-meld}, IEMOCAP~\citep{busso2008iemocap}, CMU-MOSI~\citep{zadeh-etal-2017-tensor}, CMU-MOSEI~\citep{bagher-zadeh-etal-2018-multimodal}, CH-SIMS~\citep{yu2020ch}, CH-SIMS v2~\citep{liu2022make}and OV-MERD+~\citep{lian2025ov}. Notably, CogEmo-40K improves MER2023 from \textbf{72.12 \(\rightarrow\) 74.01} (\textbf{+1.89} points) and MELD from \textbf{54.69 \(\rightarrow\) 57.35} (\textbf{+2.66} points), demonstrating that structured cognitive appraisal reasoning provides more transferable supervision than free-form affective descriptions. On the challenging CMU-MOSEI dataset, CogEmo-40K reaches \textbf{69.75} versus \textbf{70.97} for MERCaption, while on CH-SIMS v2 it trails by only \textbf{0.12} points. Overall, these results confirm that explicitly modeling the perception-to-appraisal pathway yields stronger cross-domain generalization.

\subsection{MERBench Evaluation Protocol}
\label{app:merbench}
We adopt MERBench~\citep{lian2025affectgpt} to evaluate the cross-domain generalization of different supervision paradigms. As summarized in \styleref{Table~\ref{tab:merbench_datasets}}, MERBench provides a unified evaluation protocol covering nine representative multimodal emotion datasets and three complementary affective tasks: fine-grained emotion recognition, basic emotion recognition, and sentiment analysis.

\begin{table}[htbp]
\centering
\caption{
\textbf{Statistics of the nine datasets in MERBench.}
MERBench covers fine-grained emotion recognition, basic emotion recognition,
and sentiment analysis.
}
\label{tab:merbench_datasets}

\small
\setlength{\tabcolsep}{6pt}
\renewcommand{\arraystretch}{1.12}

\begin{tabular}{llcc}
\toprule
\textbf{Task} & \textbf{Dataset} & \textbf{\# Samples} & \textbf{\# Classes} \\
\midrule

Fine-grained
& OV-MERD+ & 532 & Unified \\

\midrule

\multirow{4}{*}{Basic}
& MER2023 & 411 & 6 \\
& MER2024 & 1,169 & 6 \\
& IEMOCAP & 1,241 & 4 \\
& MELD & 2,610 & 7 \\

\midrule

\multirow{4}{*}{Sentiment}
& CMU-MOSI & 686 & 3 \\
& CMU-MOSEI & 4,659 & 3 \\
& CH-SIMS & 457 & 3 \\
& CH-SIMS v2 & 1,034 & 3 \\

\bottomrule
\end{tabular}
\end{table}

\paragraph{Fine-grained emotion recognition.}
This setting evaluates open-vocabulary emotion understanding using \textbf{OV-MERD+}~\citep{lian2025ov}, where emotion predictions are not restricted to a fixed closed-set taxonomy. Compared with conventional categorical emotion datasets, OV-MERD+ contains more diverse and fine-grained affective expressions, making it suitable for evaluating whether learned representations can generalize to open-ended emotion semantics.

\paragraph{Basic emotion recognition.}
This setting includes \textbf{MER2023}~\citep{lian2023mer}, \textbf{MER2024}~\citep{lian2024mer}, \textbf{IEMOCAP}~\citep{busso2008iemocap}, and \textbf{MELD}~\citep{poria-etal-2019-meld}. MER2023 and MER2024 are large-scale Chinese multimodal emotion datasets collected primarily from movies and television programs. IEMOCAP contains acted dyadic interactions and is evaluated under the commonly used four-class setting, while MELD consists of multi-party conversational clips from television dialogues and contains seven basic emotion categories.

\paragraph{Sentiment analysis.}
This setting includes \textbf{CMU-MOSI}~\citep{zadeh-etal-2017-tensor}, \textbf{CMU-MOSEI}~\citep{bagher-zadeh-etal-2018-multimodal}, \textbf{CH-SIMS}~\citep{yu2020ch}, and \textbf{CH-SIMS v2}~\citep{liu2022make}. CMU-MOSI~\citep{zadeh-etal-2017-tensor} and CMU-MOSEI contain multimodal opinion videos with sentiment annotations, with CMU-MOSEI providing a substantially larger-scale extension. CH-SIMS and CH-SIMS v2 consist of Chinese multimodal sentiment samples and provide complementary evaluation in a different language and cultural context.

Following the original MERBench protocol, task-specific metrics are used for the three settings. Fine-grained emotion recognition adopts a set-based evaluation after normalizing semantically equivalent emotion expressions; basic emotion recognition evaluates whether the predicted emotion matches the ground-truth basic category; and sentiment analysis reports class-imbalance-aware classification performance after mapping continuous sentiment annotations into discrete polarity categories. Although MERBench is built on conventional emotion recognition and sentiment datasets, the evaluated models produce free-form emotion analyses that are mapped to benchmark label spaces for scoring. Our perception-to-appraisal supervision further structures these analyses through cognitive dimensions, allowing us to test whether cognition-grounded emotion understanding transfers to standard downstream benchmarks.

All compared models in our cross-domain experiments follow the same MERBench preprocessing, prompting, and metric computation protocol to ensure consistent evaluation.

\subsection{Case Study: Qualitative Comparison of Cognitive Appraisal Reasoning}
\label{app:case_study}

To complement the quantitative results in \styleref{Table~\ref{tab:cogemo_bench}}, we present a qualitative case study comparing the reasoning trajectories of our CogEmo-MoE with the best-performing baseline (GPT-5.2). We selected three samples from CogEmo-Bench that highlight typical failure modes of existing MLLMs: (1) emotion confusion between contempt and anger, (2) surface-level textual misleading, and (3) over-interpretation of neutral contexts. For each sample, we show the ground-truth cognitive appraisal reasoning, the output of our model, and the output of GPT-5.2.

\subsubsection{Case 1: Distinguishing Contempt from Anger}

\begin{figure}[H]
    \centering
    \begin{subfigure}{0.32\linewidth}
        \includegraphics[width=\linewidth]{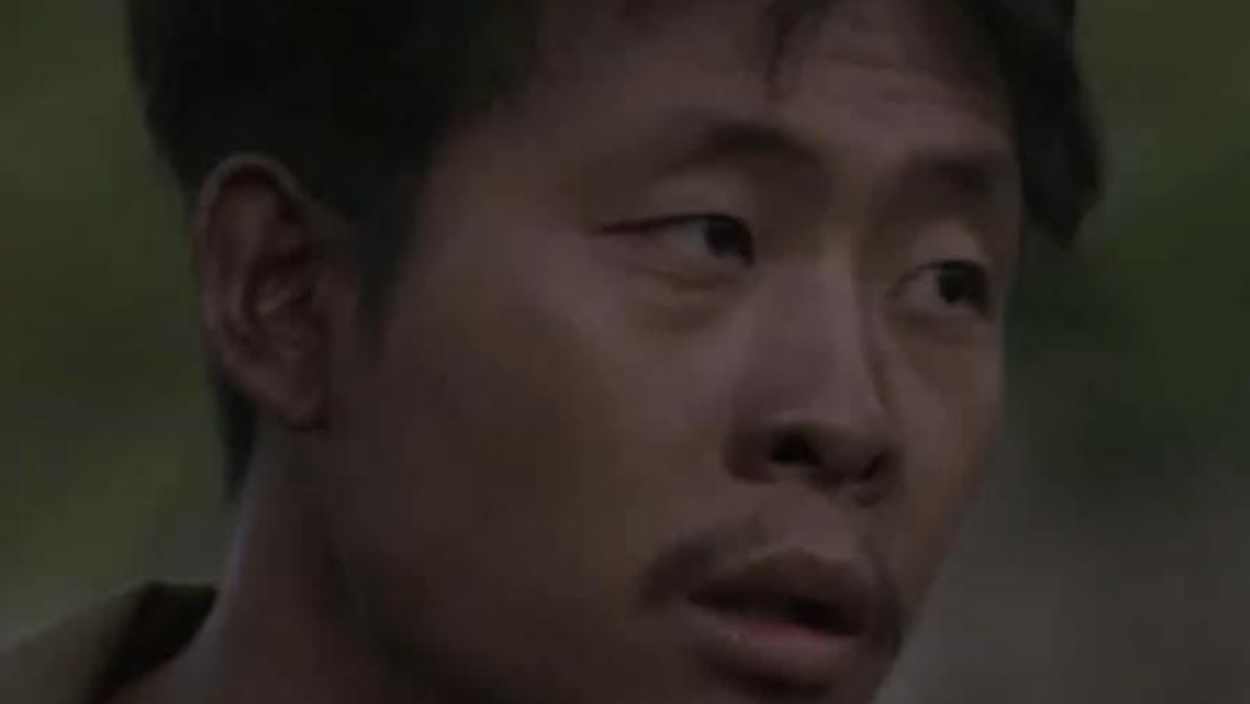}
        \caption{Frame 1}
        \label{fig:case1_frame1}
    \end{subfigure}
    \hfill
    \begin{subfigure}{0.32\linewidth}
        \includegraphics[width=\linewidth]{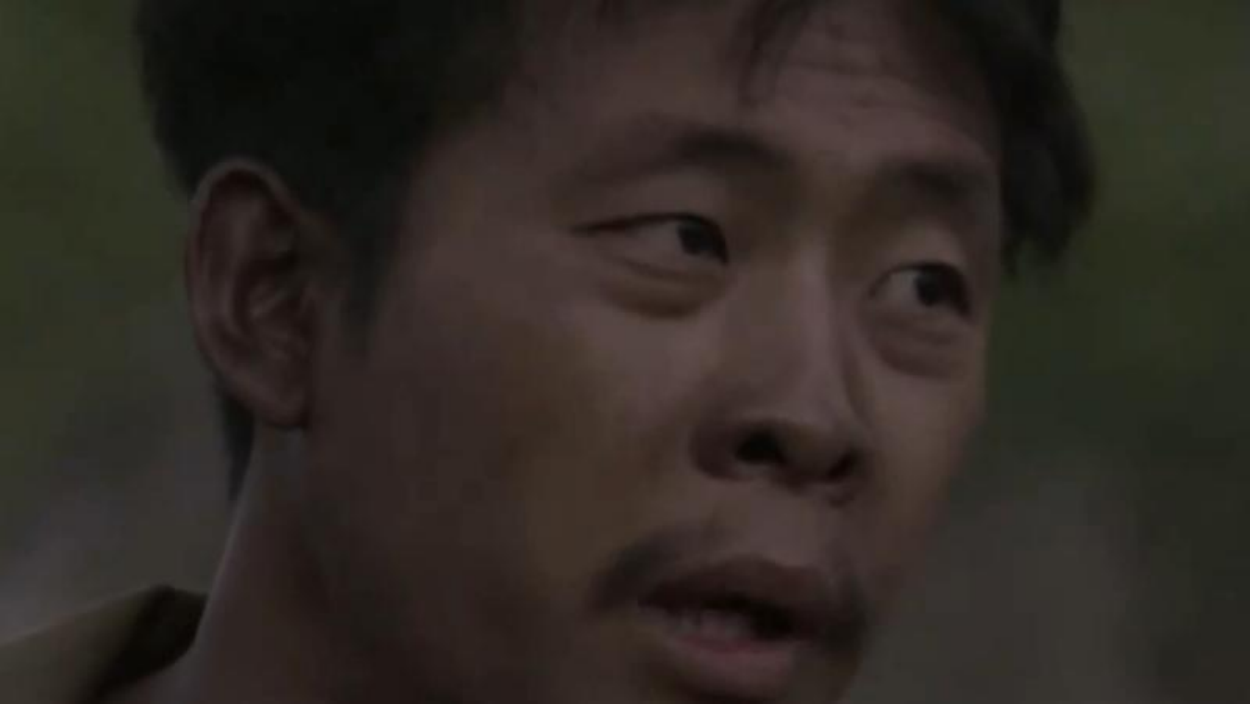}
        \caption{Frame 2}
        \label{fig:case1_frame2}
    \end{subfigure}
    \hfill
    \begin{subfigure}{0.32\linewidth}
        \includegraphics[width=\linewidth]{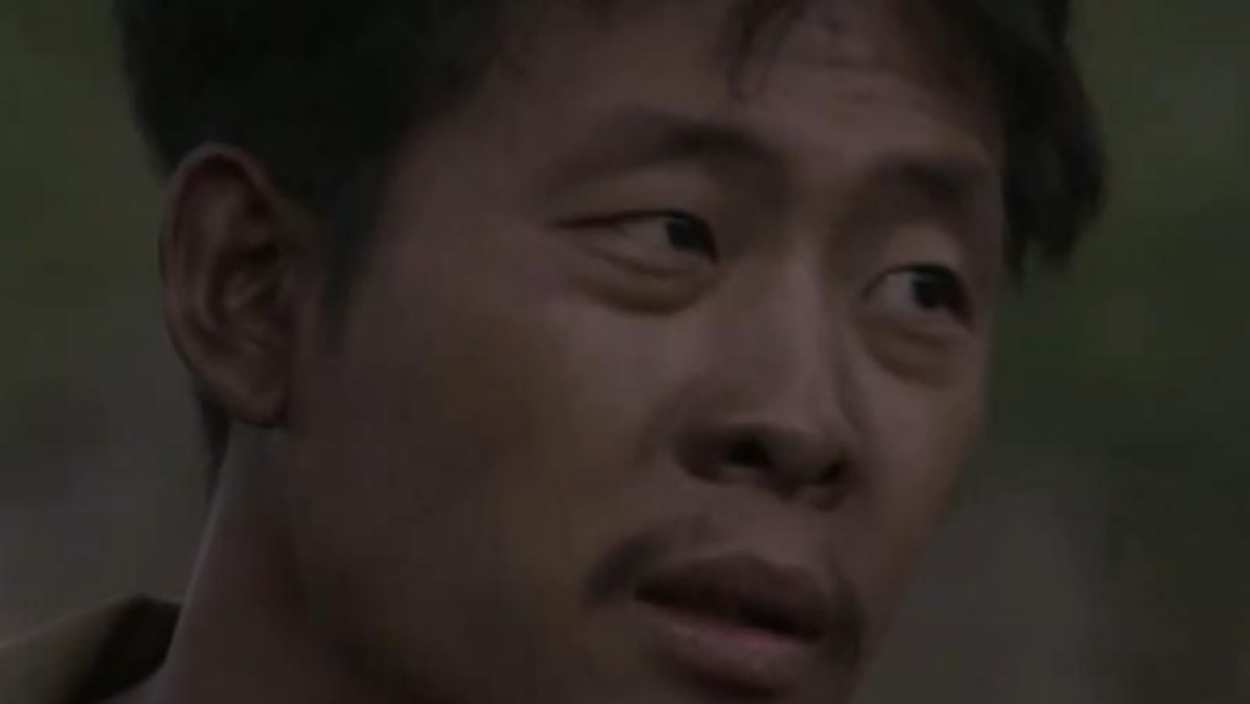}
        \caption{Frame 3}
        \label{fig:case1_frame3}
    \end{subfigure}
    \caption{Selected video frames for Case 1 (Contempt vs.\ Anger). The frames show the controlled downward gaze, slightly downturned mouth corners, and stable head pose characteristic of contempt, as opposed to high-arousal anger cues.}
    \label{fig:case1_frames}
\end{figure}

\textbf{Ground Truth Emotion:} Contempt \\
\textbf{Context:} A man is criticized as ``the weakest link'' in a military confrontation.

\paragraph{Ground-Truth Cognitive Appraisal:}
\begin{quote}
\small
``The dismissive verbal content calling the others merely `three hundred defeated soldiers,' together with the downturned lip corners and steady off-camera address, suggests a devaluing, unfavorable evaluation of the target; meanwhile, his clear articulation, moderate voice intensity, stable gaze, and relaxed jaw imply he feels able to handle the situation and speak from a position of control; furthermore, because the criticism is directed outward at the off-camera group through the insulting label `defeated soldiers,' the cause of concern is assigned to others rather than himself. Overall, the pattern is one of negative evaluation, confident control, blame toward others, present-focused engagement, and restrained activation, therefore the most likely emotion is contempt.''
\end{quote}

\paragraph{Our Model (CogEmo-MoE) Output:}
\begin{quote}
\small
``The man's steady slight downward gaze, mildly downturned mouth corners, and the critical verbal content (`you are the weakest link') convey an unfavorable evaluation of the target; meanwhile, his controlled posture, clear articulation, and stable low-mobility face with moderate vocal intensity suggest he remains composed enough to address the issue directly rather than being overwhelmed; furthermore, because the utterance explicitly points to `you' as the source of the problem and his facial expression stays directed outward rather than self-reproachful, responsibility is assigned externally. Overall, the pattern is one of controlled negative evaluation with external blame, present-focused criticism, low arousal, and direct engagement, therefore the most likely emotion is contempt.''
\end{quote}
\textbf{Prediction:} Contempt $\checkmark$

\paragraph{Baseline Model (GPT-5.2) Output:}
\begin{quote}
\small
``Across the sampled frames, the person appears tense and increasingly confrontational. His brows are drawn and his gaze is fixed off to the side as if reacting to someone nearby; his mouth stays slightly open as though speaking defensively. The final frame shows a sharp escalation: eyes widened, jaw set, and teeth visible in what looks like a raised voice or shouted remark—classic cues of agitation and aggression... Together, the facial tension, sudden wide-eyed glare, and the insulting/critical text point to an angry response rather than sadness or fear.''
\end{quote}
\textbf{Prediction:} Anger $\times$ (Ground Truth: Contempt)

\paragraph{Analysis:}
The baseline model interprets the raised voice and confrontational facial cues as anger, focusing on observable affective signals of agitation. In contrast, our model explicitly reasons about \textit{Accountability} (``responsibility is assigned externally'') and \textit{Coping} (``remains composed enough to address the issue directly''), which are the key appraisal dimensions distinguishing contempt (characterized by moral superiority and external blame with low arousal) from anger (characterized by high arousal and goal obstruction). This demonstrates that cognitive appraisal reasoning enables our model to differentiate emotionally similar but cognitively distinct negative states.

\subsubsection{Case 2: Overcoming Surface-Level Textual Misleading}

\begin{figure}[H]
    \centering
    \begin{subfigure}{0.32\linewidth}
        \includegraphics[width=\linewidth]{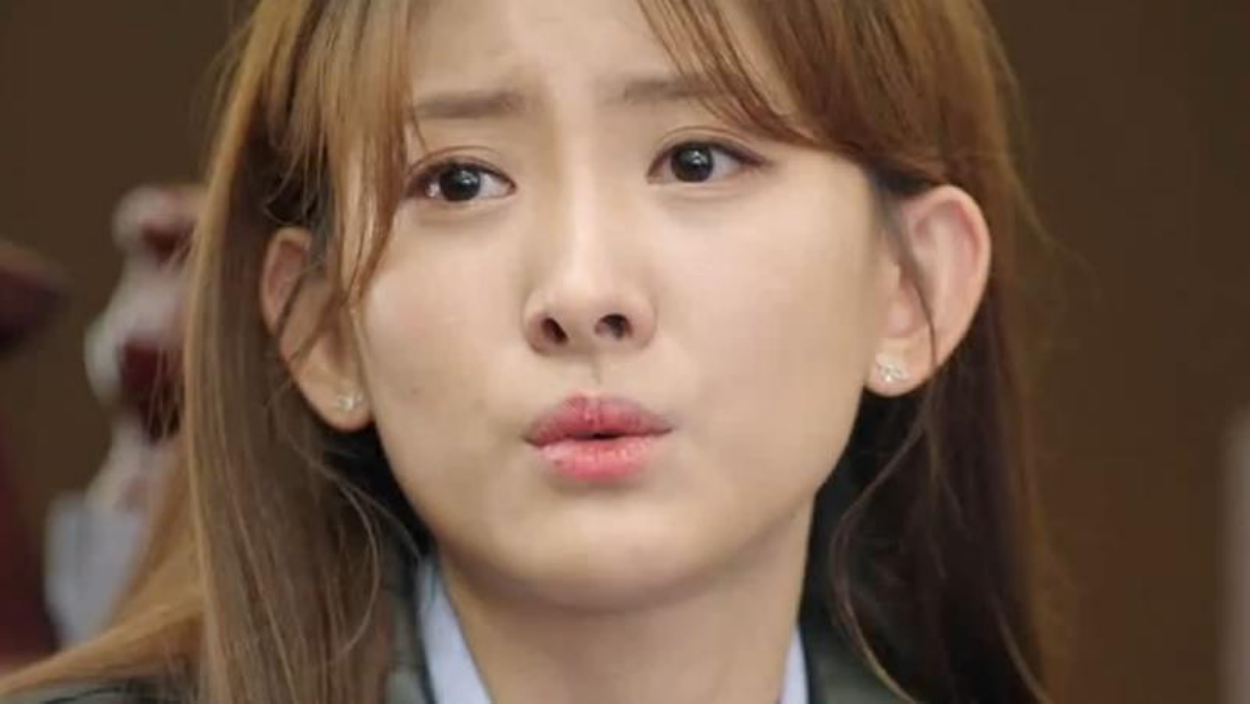}
        \caption{Frame 1}
        \label{fig:case2_frame1}
    \end{subfigure}
    \hfill
    \begin{subfigure}{0.32\linewidth}
        \includegraphics[width=\linewidth]{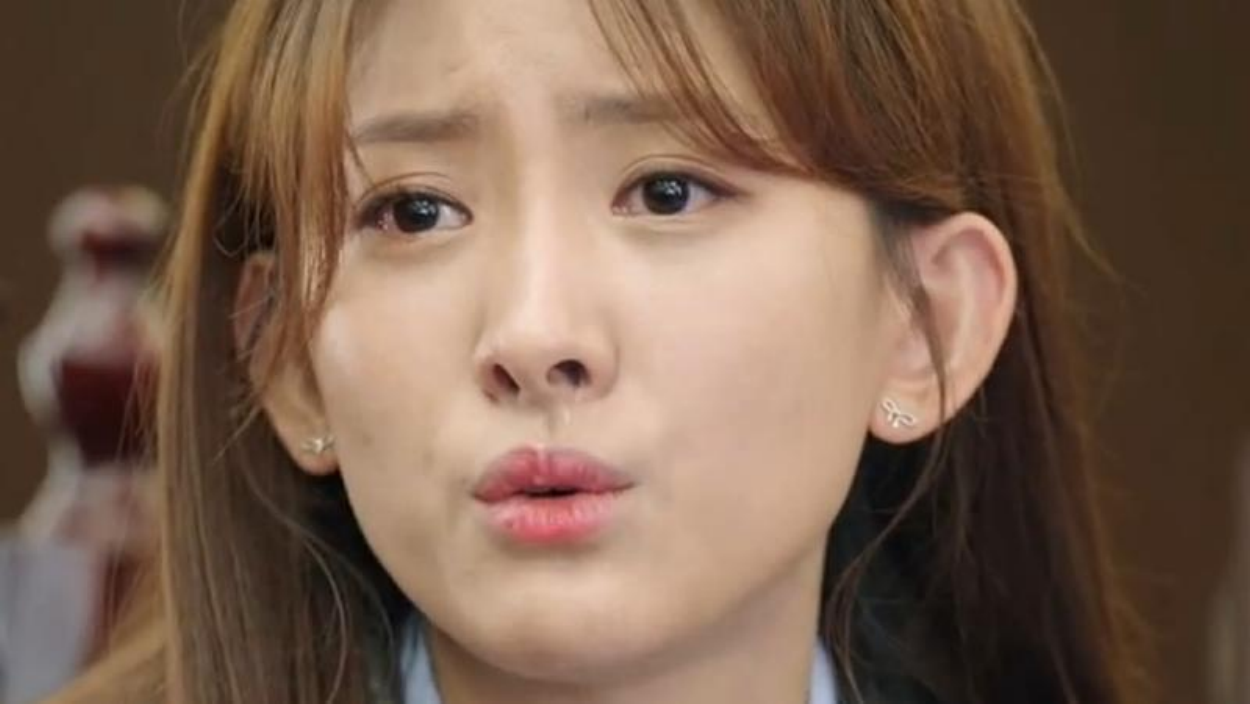}
        \caption{Frame 2}
        \label{fig:case2_frame2}
    \end{subfigure}
    \hfill
    \begin{subfigure}{0.32\linewidth}
        \includegraphics[width=\linewidth]{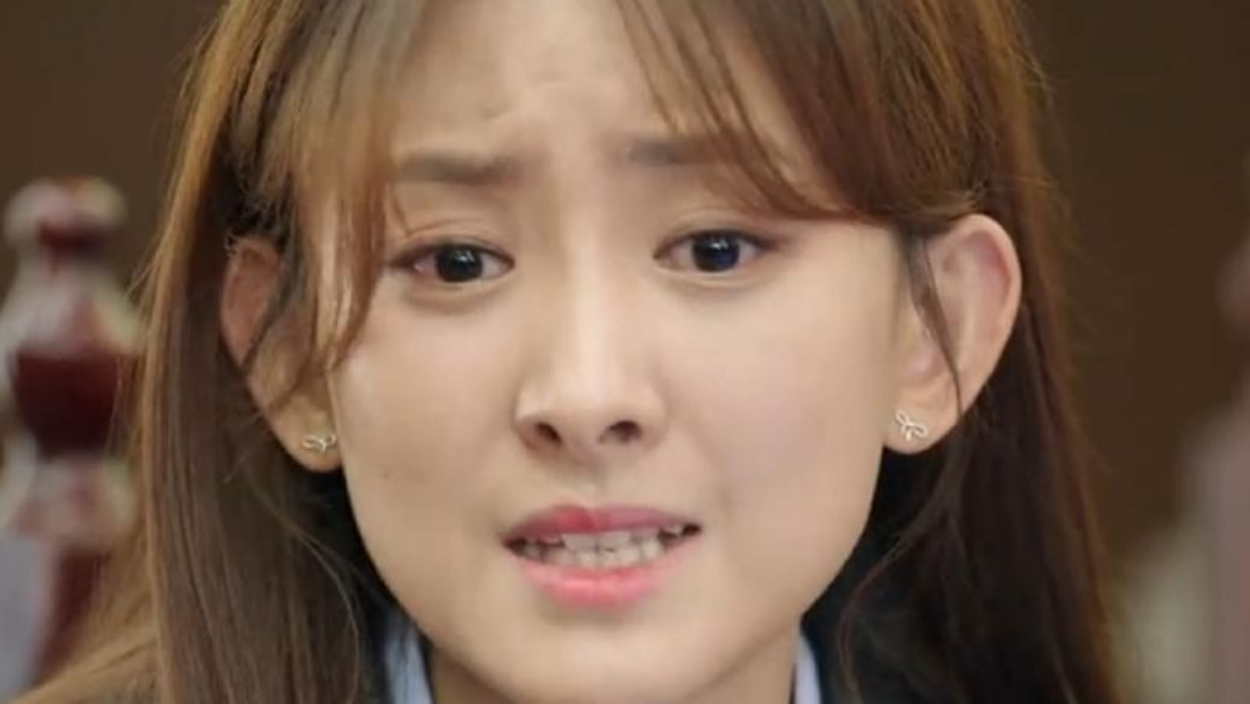}
        \caption{Frame 3}
        \label{fig:case2_frame3}
    \end{subfigure}
    \caption{Selected video frames for Case 2 (Fear vs.\ Sadness). The frames illustrate the tightened jaw, narrowed eyes, and tense mouth while recounting being comforted, indicating a lingering threat appraisal rather than past-oriented sadness.}
    \label{fig:case2_frames}
\end{figure}

\textbf{Ground Truth Emotion:} Fear \\
\textbf{Context:} A woman testifies in court, recounting that someone comforted her.

\paragraph{Ground-Truth Cognitive Appraisal:}
\begin{quote}
\small
``Her tightened jaw, narrowed eyes, raised yet unstable brows, and downturned but tense mouth while recounting that she felt okay because another person was there in a formal hearing-like setting suggest an unpleasant and threatening evaluation of the situation; meanwhile, although she continues speaking clearly and remains physically composed, the growing facial tension and reliance in her words on another person's reassurance imply only limited personal control over the threatening situation; furthermore, because her statement frames safety as depending on someone else's presence and comfort rather than her own action, the source of regulation is centered outside herself... Overall, the pattern is of negatively evaluated threat, limited control, externally centered responsibility, anticipation of danger, high activation, and a withdrawal-oriented response, therefore the emotion is fear.''
\end{quote}

\paragraph{Our Model (CogEmo-MoE) Output:}
\begin{quote}
\small
``The speaker's statement that someone comforted her, together with the otherwise subdued and serious interaction, points to a negatively toned situation rather than a pleasant one; meanwhile, the relaxed face, steady gaze, moderate voice, and lack of forceful movement suggest the person is not actively exerting strong control over the situation; furthermore, the spoken content centers on another person's comforting behavior as the relevant cause, which places causation outside the speaker... Overall, the pattern is of a negative, externally caused, ongoing situation with limited coping, low arousal, and withdrawal, therefore the emotion is fear.''
\end{quote}
\textbf{Prediction:} Fear $\checkmark$

\paragraph{Baseline Model (GPT-5.2) Output:}
\begin{quote}
\small
``The woman appears emotionally distressed and vulnerable. Her furrowed brows, raised inner eyebrows, tense eyes, pursed/downturned lips, and lowered gaze suggest worry and lingering sadness rather than happiness or neutrality. The subtitles indicate that Mu Zi comforted her and that she felt things would be okay because Mu Zi was there, implying she had previously been upset or anxious and is now only partially reassured. Her expression remains strained while speaking, suggesting emotional pain and concern beneath the relief... Most likely emotion: sadness.''
\end{quote}
\textbf{Prediction:} Sadness $\times$ (Ground Truth: Fear)

\paragraph{Analysis:}
The baseline model appears to rely on the textual cues ``comforted'' and ``felt better'' and predicts sadness, interpreting the situation as grief or loss. Our model instead describes an ongoing situation and limited coping, which are more consistent with the fear label in this example. The case illustrates how appraisal-based descriptions can help distinguish interpretations when the textual cues are ambiguous.

\subsubsection{Case 3: Correctly Identifying Neutral Affect}

\begin{figure}[H]
    \centering
    \begin{subfigure}{0.32\linewidth}
        \includegraphics[width=\linewidth]{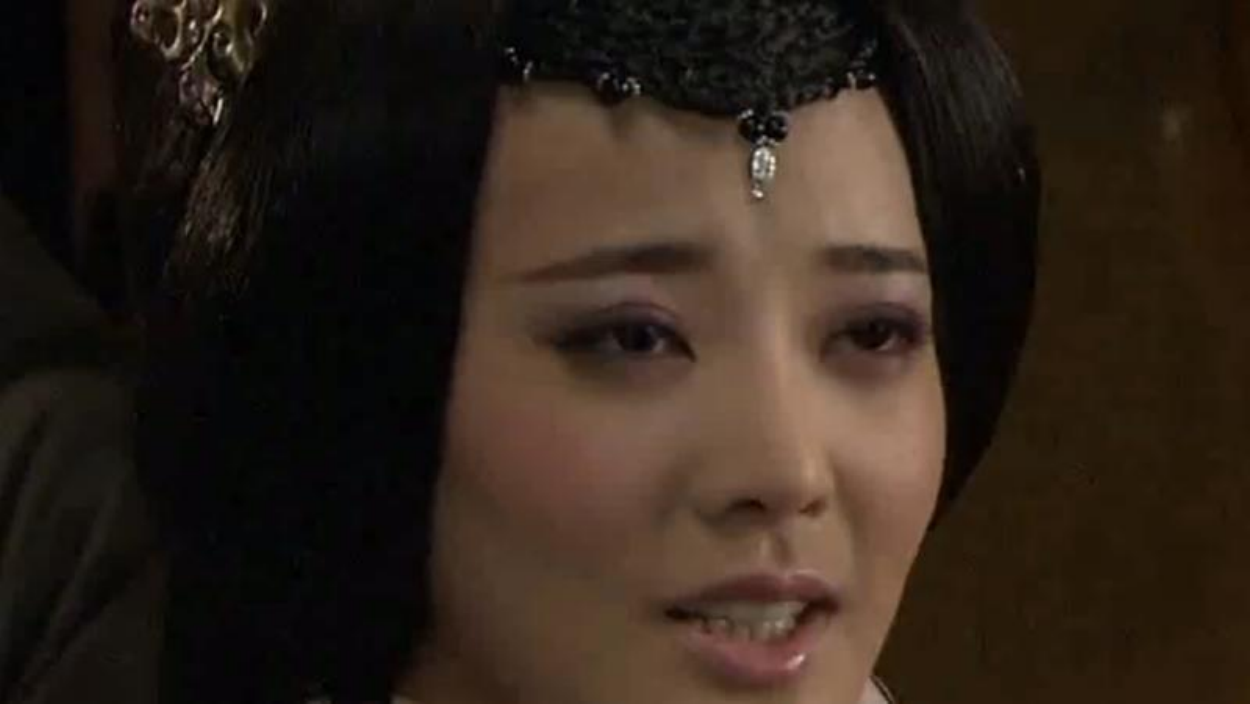}
        \caption{Frame 1}
        \label{fig:case3_frame1}
    \end{subfigure}
    \hfill
    \begin{subfigure}{0.32\linewidth}
        \includegraphics[width=\linewidth]{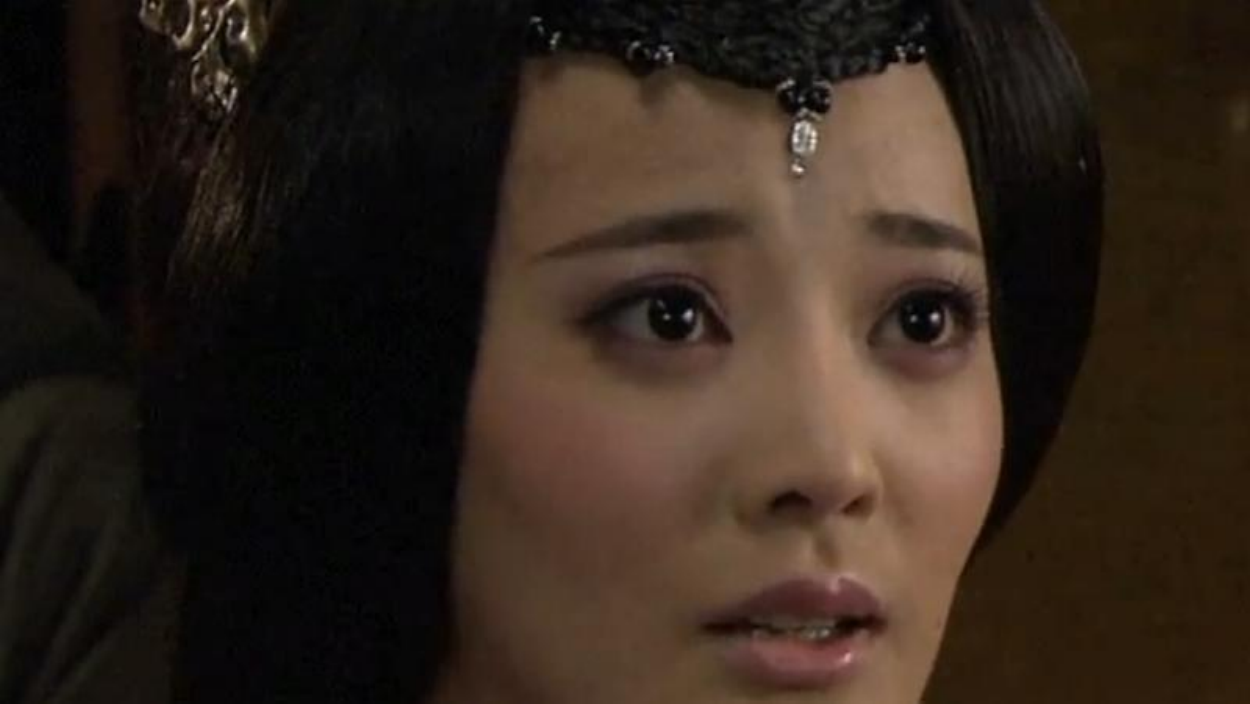}
        \caption{Frame 2}
        \label{fig:case3_frame2}
    \end{subfigure}
    \hfill
    \begin{subfigure}{0.32\linewidth}
        \includegraphics[width=\linewidth]{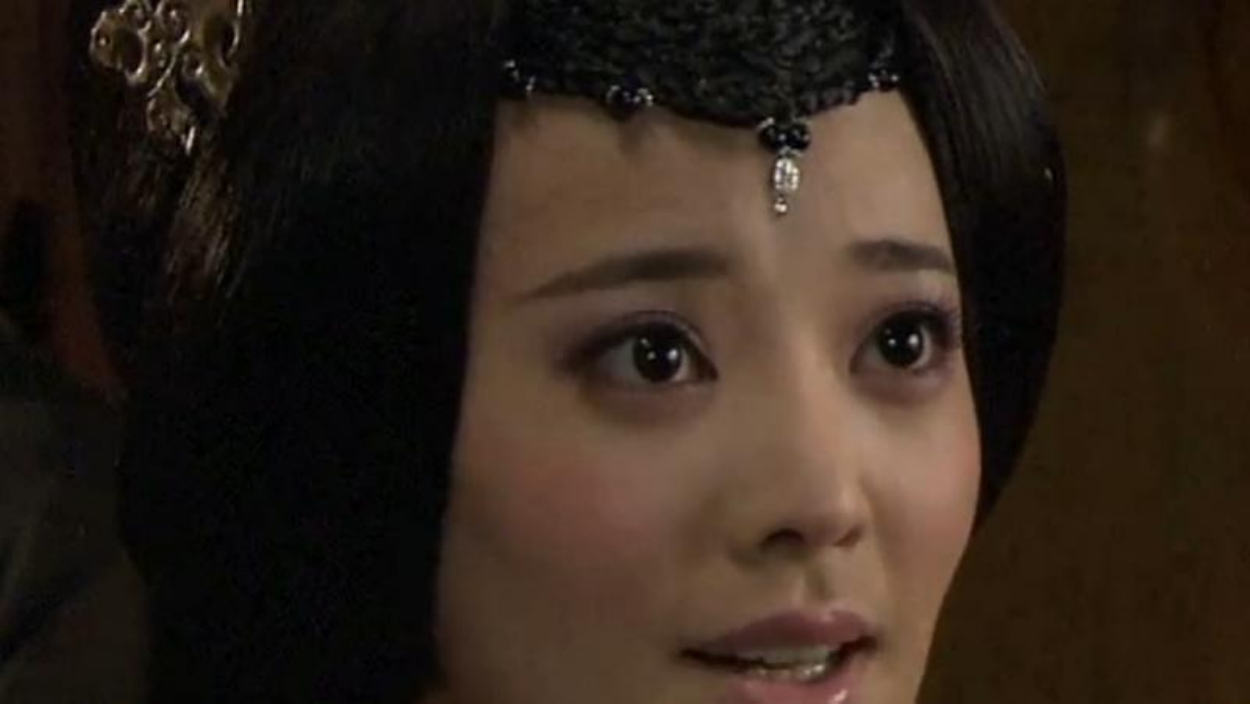}
        \caption{Frame 3}
        \label{fig:case3_frame3}
    \end{subfigure}
    \caption{Selected video frames for Case 3 (Neutral vs.\ Anxiety). The woman's relaxed facial expression, stable gaze, and slight smile indicate low arousal and adequate coping, supporting a neutral appraisal rather than anxiety.}
    \label{fig:case3_frames}
\end{figure}

\textbf{Ground Truth Emotion:} Neutral \\
\textbf{Context:} A woman politely apologizes and explains that someone has offended a noble.

\paragraph{Ground-Truth Cognitive Appraisal:}
\begin{quote}
\small
``The woman's polite wording asking for an audience, along with her relaxed face, slight mouth upturn, and steady moderate speech toward the man, suggests the interaction is not strongly adverse and may even be socially formal; her clear articulation, stable head pose, direct gaze, and controlled vocal delivery imply she feels able to manage the situation and speak deliberately; meanwhile, because she explicitly says she is sorry and that Wang Shang has offended Mi Yuan, the source of the issue is framed as lying with external parties rather than herself... Overall, the pattern is mildly positive, controlled, externally attributed, present-focused, low-arousal, and socially engaging, therefore the most likely emotion is neutral.''
\end{quote}

\paragraph{Our Model (CogEmo-MoE) Output:}
\begin{quote}
\small
``The woman's polite wording asking for an audience, along with her relaxed face, slight mouth upturn, and steady moderate speech toward the man, suggests the interaction is not strongly adverse and may even be socially formal; her clear articulation, stable head pose, direct gaze, and controlled vocal delivery imply she feels able to manage the situation... Overall, the pattern is mildly positive, controlled, externally attributed, present-focused, low-arousal, and socially engaging, therefore the most likely emotion is neutral.''
\end{quote}
\textbf{Prediction:} Neutral $\checkmark$

\paragraph{Baseline Model (GPT-5.2) Output:}
\begin{quote}
\small
``The woman appears emotionally distressed and concerned. Her raised, knitted eyebrows, slightly widened eyes, tense mouth, and pleading gaze toward the man suggest worry about the interpersonal conflict. The text indicates that Wang Shang has offended Mi Yuan, and she apologizes (`Sorry to bother you'), implying she may feel anxious about the consequences... Most likely emotion: anxiety.''
\end{quote}
\textbf{Prediction:} Anxiety $\times$ (Ground Truth: Neutral)

\paragraph{Analysis:}
The baseline model over-interprets the apologetic and formal social context as emotional distress, projecting anxiety onto a neutral interaction. Our model correctly recognizes that the woman's \textit{Coping} (``able to manage the situation'') and \textit{Arousal} (low activation, ``relaxed face'') do not support an anxiety interpretation. This case illustrates how appraisal reasoning may help the model avoid assigning anxiety to a neutral conversational exchange.

\subsubsection{Summary of Case Study}

Across these three representative cases, we observe consistent patterns in how our model outperforms the baseline:

\begin{table}[H]
\centering
\caption{Appraisal dimensions correctly identified by CogEmo-MoE in each case study.}
\label{tab:case_study_summary}
\setlength{\tabcolsep}{6pt}
\renewcommand{\arraystretch}{1.15}
\begin{tabular}{lccc}
\toprule
\textbf{Dimension} & \textbf{Case 1} & \textbf{Case 2} & \textbf{Case 3} \\
 & (Contempt vs. Anger) & (Fear vs. Sadness) & (Neutral vs. Anxiety) \\
\midrule
Goal Congruence & Negative evaluation & Negative threat & Not strongly adverse \\
Coping Potential & High (controlled) & Low (limited) & High (able) \\
Accountability & External blame & External cause & External parties \\
Future Expectancy & Present-focused & Anticipatory threat & Present-focused \\
Arousal & Low (restrained) & High (activation) & Low (relaxed) \\
Action Tendency & Confrontational & Withdrawal & Socially engaging \\
\bottomrule
\end{tabular}
\end{table}

These qualitative examples illustrate that cognitive appraisal reasoning provides critical discriminative information that is not accessible through surface-level affective cue matching alone. By explicitly modeling how individuals evaluate situations along dimensions such as accountability, coping potential, and future expectancy, our model achieves more reliable emotion understanding, particularly in cases involving subtle emotional distinctions, misleading textual content, or contextually appropriate neutral affect.

We acknowledge that these three cases are qualitatively selected for illustrative purposes; the statistical superiority of our model is rigorously validated in the main experiments (\styleref{Table~\ref{tab:cogemo_bench}}).

\end{document}